\documentclass[11pt]{article}
\usepackage[utf8]{inputenc}
\usepackage{array}
\usepackage[T1]{fontenc}
\usepackage{tabu}
\usepackage{mathtools}
\usepackage{bm}
\usepackage{authblk}
\usepackage{amsfonts}
\usepackage{color}
\usepackage{tabu}
\usepackage{colortbl}

\usepackage{calrsfs}
\DeclareMathAlphabet{\pazocal}{OMS}{zplm}{m}{n}

\newcommand{\thickhline}{\noalign{\hrule height 2pt}}
\newcolumntype{+}{!{\vrule width 2pt}}
\definecolor{lightgray}{gray}{0.9}

\title{Optimisation of the coalescent hyperbolic embedding of complex networks}
\date{}

\author[1]{Bianka Kovács}
\author[1,2,3,*]{Gergely Palla}
\affil[1]{Dept.\ of Biological Physics, Eötvös Lor{\'a}nd University, H-1117 Budapest, P{\'a}zm{\'a}ny P.\ stny.\ 1/A, Hungary}
\affil[2]{MTA-ELTE Statistical and Biological Physics Research Group, H-1117 Budapest, P{\'a}zm{\'a}ny P.\ stny.\ 1/A, Hungary}
\affil[3]{Digital Health and Data Utilisation Team, Health Services Management Training Centre,Faculty of Health and Public Administration, Semmelweis University, Budapest, Hungary}

\affil[*]{pallag@hal.elte.hu}

\begin{document}
\maketitle

\begin{abstract}
Several observations indicate the existence of a latent hyperbolic space behind real networks that makes their structure very intuitive in the sense that the probability for a connection is decreasing with the hyperbolic distance between the nodes. A remarkable network model generating random graphs along this line is the popularity-similarity optimisation (PSO) model, offering a scale-free degree distribution, high clustering and the small world property at the same time. These results provide a strong motivation for the development of hyperbolic embedding algorithms, that tackle the problem of finding the optimal hyperbolic coordinates of the nodes based on the network structure. A very promising recent approach for hyperbolic embedding is provided by the noncentered minimum curvilinear embedding (ncMCE) method, belonging to the family of coalescent embedding algorithms. This approach offers a high quality embedding at a low running time. In the present work we propose a further optimisation of the angular coordinates in this framework that seems to reduce the logarithmic loss and increase the greedy routing score of the embedding compared to the original version, thereby adding an extra improvement to the quality of the inferred hyperbolic coordinates. 
\end{abstract}
%\begin{document}

%\flushbottom
%\maketitle
% * <john.hammersley@gmail.com> 2015-02-09T12:07:31.197Z:
%
%  Click the title above to edit the author information and abstract
%
%\thispagestyle{empty}

%\noindent Please note: Abbreviations should be introduced at the first mention in the main text – no abbreviations lists. Suggested structure of main text (not enforced) is provided below.

\section{Introduction}
Network theory has become ubiquitous in the study of complex systems composed of many interacting units \cite{Laci_revmod,Dorog_book,Newman_Barabasi_Watts}. Over the last two decades, the overwhelming number of studies using this approach in systems ranging from metabolic interactions to the level of the global economy have shown that the statistical analysis of the underlying graph structure can highlight non-trivial properties and reveal previously unseen relations \cite{Laci_revmod,Dorog_book,Newman_Barabasi_Watts,Jari_Holme_book,Vespignani_book}. Probably the most important universal features of networks representing real systems are the small world property \cite{Milgram_small_world,Kochen_book}, the high clustering coefficient \cite{Watts-Strogatz} and the scale-free degree distribution \cite{Faloutsos,Laci_science}. On the modelling ground, a large number of network models were proposed for capturing one (or several) of these properties in a simple mathematical framework, and a quite notable example among these is provided by the PSO model \cite{PSO}, which reproduces all three properties simultaneously in a natural manner. In this approach the nodes are placed one by one on the native disk representation\cite{hyperGeomBasics} of the 2D hyperbolic plane with a logarithmically increasing radial coordinate and a random angular coordinate, and links are drawn with probabilities determined by the hyperbolic distance between the node pairs. In vague terms, the degree of nodes is determined by their radial coordinate (lower distance from the origin corresponds to larger degree), and the angular proximity of the nodes can be interpreted as a sort of similarity, where more similar nodes have a higher probability to be connected. 
 
 % A saját rejtett paraméteres cikkunkre kéne hivatkozni!! 
The idea that hidden metric spaces can play an important role in the structure of complex networks first arose in a study focusing on the self-similarity of scale-free networks \cite{Boguna_2008_PRL}. This was followed by reports showing the signs of hidden geometric spaces behind protein interaction networks \cite{Higham_geom_protein_2008,Kuchaiev_geom_protein_2009}, the Internet \cite{Boguna_2009_nat_phys, Boguna_Krioukov_Internet_2010,Jonckhere_Internet_2011,Bianconi_internet_2015,Chepoi_Internet_2017}, brain networks \cite{Cannistraci_brain_2013,Tadic_brain_2018}, or the world trade network \cite{Boguna_trade_net_2016}, also revealing important connections between the the navigability of networks and hyperbolic spaces \cite{Boguna_2009_nat_phys,Gulyas_natcoms}. In parallel, practical tools for generating hyperbolic networks \cite{von_Looz_generate} and methods for measuring the hyperbolicity of networks were also proposed \cite{Kennedy_measure_hyper,Borassi_measury_hyper}. In the recent years the geometric nature of weights \cite{Boguna_geometric_weights_ncoms} and clustering \cite{Candellero_clust_hyp_geom,Krioukov_clust_hyp_geom} was revealed, and further variants of the original PSO model were proposed for generating random hyperbolic networks with communities \cite{Zuev_soft_hyper_coms,nPSO_New_J_Phys}. 
 %where the concept of greedy routing was introduced. When walking on the network according to this process, we always choose the neighbour of the current node that has the closest geometrical position to the target. 
 
The very notable advancements in the hidden metric space related research provide a strong motivation for the development of hyperbolic embedding techniques \cite{HyperMap,Alanis-Lobato_LE_embedding,coalescentEmbedding,Alanis-Lobat_liekly_LE_emb,Boguna_embedding_2019}, that tackle the problem of inferring plausible coordinates for the nodes based on the network structure. One of the first methods pointing in this direction was HyperMap \cite{HyperMap}, which optimises a logarithmic loss function obtained from the assumption that the network was generated according to a generalised version of the PSO model (referred to as the E-PSO model). In contrast, in Ref.\cite{Alanis-Lobato_LE_embedding} an embedding based on a non-linear dimension reduction of the Laplacian matrix was proposed. Along a similar line, a whole family of embedding algorithms were studied in Ref.\cite{coalescentEmbedding}, using different pre-weighted matrices encapsulating the network structure and multiple unsupervised dimension reduction techniques from machine learning. In this framework, after the dimension reduction the nodes are organised on a circular or quasilinear manifold from which the angular coordinates in the 2D hyperbolic plane can be obtained in a simple manner, whereas the radial coordinates are inferred based on the node degrees. The rationale behind such an approach is that for networks that are actually generated in a hyperbolic manner, the angular order of the nodes is preserved along the obtained low dimensional manifold. This phenomenon is referred to as 'angular coalescence', and thus these methods are called coalescent embedding algorithms \cite{coalescentEmbedding}. A combination of the Laplacian embedding and the likelihood optimisation based on the E-PSO model was proposed in Ref.\cite{Alanis-Lobat_liekly_LE_emb}, and in a recent work the approach named Mercator was introduced \cite{Boguna_embedding_2019}, where the Laplacian embedding is incorporated with optimisation with respect to the so-called $\mathbb{S}^1/\mathbb{H}^2$ model \cite{Boguna_2008_PRL}.
 
In the present paper we propose an embedding algorithm combining a coalescent approach with likelihood optimisation based on the E-PSO model. One of the best performing dimension reduction techniques in Ref.\cite{coalescentEmbedding} was corresponding to the non-centered minimum curvilinear embedding (ncMCE) \cite{ncMCE}, which also provides the starting point of our method. However, after obtaining the initial node coordinates based on ncMCE, we also apply an angular optimisation of the coordinates using a logarithmic loss function originating from the E-PSO model. We test the proposed approach on both synthetic and real network data, and compare the results with the outcome of HyperMap, the orignal ncMCE coalescent embedding and Mercator in terms of the achieved logarithmic loss and the greedy routing score (which is a model-free quality measure of the embeddings). 
 
% This approach is tested on both synthetic and real network data. Based on comparison with HyperMap, the orignal ncMCE coalescent embedding and Mercator, our method is competitive with current state-of-the-art algorithms both in terms of the achieved loss and the greedy rooting score (corresponding to a model-free quality measure of the embedding methods). In addition, for real networks with inherent community structure, the angular optimisation seems to provide a natural separation between the different modules, resulting in a layout showing the community structure in more transparent manner compared to the original coalescent embedding. 

\section{Preliminaries and algorithm description} 
In the following, we briefly describe the necessary preliminaries together with our angular optimisation algorithm. Since the optimisation uses a logarithmic loss function based on the E-PSO model, we begin with the PSO and E-PSO models, together with the loss function and a short description of the HyperMap embedding algorithm. These are followed by the summary of the coalescent embedding algorithm ncMCE and the proposed optimisation of the angular coordinates.

\subsection{The E-PSO model and HyperMap}

The basic idea of the PSO model is to place nodes on the native disk representation of the hyperbolic plane with increasing radial coordinates and random angular coordinates, and connect the node pairs with a linking probability depending on their hyperbolic distances. The parameters of the model are the curvature of the hyperbolic plane $K<0$ parametrised by $\zeta = \sqrt{-K}$, the total number of nodes $N$, the average degree $\left< k\right>$ parametrised by $m=\left< k\right>/2$, the popularity fading parameter $\beta\in(0,1]$ controlling the outward drift of the nodes, and the  'temperature' $T\in[0,1)$ regulating the average clustering coefficient of the generated network. Initially the network is empty, and the nodes are placed on the hyperbolic disk in an iterative manner according to the following rules:
\begin{enumerate}
    \item At iteration $i$ the new node $i$ appears with the radial coordinate $r_{ii} = \frac{2}{\zeta}\ln i$ and a uniformly random angular coordinate $\theta_i\in[0,2\pi)$. (The double indexing of the radial coordinate is for a simple book keeping of the position during the outward drift specified in the next rule).
    \item The radial coordinates of all previous nodes $j<i$ are increased as $r_{ji} = \beta r_{jj}+(1-\beta)r_{ii}$. (Thus, the first index of the node position refers to the moment of birth, whereas the second index corresponds to the actual time step). This repeated outward shift in the node positions is usually referred to as 'popularity fading', since nodes closer to the origin of the hyperbolic disk are close (in the hyperbolic sense) to a higher number of other nodes compared to nodes on the periphery. 
    \item The new node $i$ is attached to the already existing nodes as follows:
    \begin{itemize}
        \item[a)] If the number of previous nodes is $m$ or smaller, then $i$ is connected to all of them.
        \item[b)] Otherwise, if $T=0$, then node $i$ is connected to the $m$ closest nodes according to the hyperbolic distance $x_{ij}$. For nodes with polar coordinates $(r_{ii},\theta_i)$ and $(r_{ji},\theta_j)$ this can be calculated from the hyperbolic law of cosines as
        \begin{equation}
            \cosh(\zeta x_{ij})  = \cosh(\zeta r_{ii})\cosh(\zeta r_{ji})-\sinh(\zeta r_{ii})\sinh(\zeta r_{ji})\cos(\Delta \theta),
            \label{eq:hyper_distance}
        \end{equation}
        where the angular difference $\Delta \theta$ is given by $\Delta \theta=\pi -\left| \pi -\left|\theta_i -\theta_j\right|\right|$.
        \item[c)] If $i>m+1$ and $T>0$, then node $i$ is connected to nodes $j<i$ with a probability depending on the hyperbolic distance $x_{ij}$ as
        \begin{equation}
            p(x_{ij})=\frac{1}{1+\mathrm{e}^{\frac{\zeta}{2T}(x_{ij}-R_i)}}, \label{eq:PSO_link_prob}
        \end{equation}
        where the cutoff distance $R_i$ is given by
        \begin{equation}
            R_i = \left\lbrace \begin{array}{ll} 
            r_{ii}-\frac{2}{\zeta}\ln\left(\frac{2T}{\sin(T\pi)}\cdot
             \frac{1-\mathrm{e}^{-\frac{\zeta}{2}(1-\beta)r_{ii}}}{m(1-\beta)}\right) & \mathrm{if}\;\; \beta < 1, \\
             r_{ii} - \frac{2}{\zeta}\ln\left(\frac{T}{\sin(T\pi)}\cdot \frac{\zeta r_{ii}}{m}\right) & \mathrm{if} \;\; \beta=1. \end{array} \right.
             \label{eq:cutoff}
        \end{equation}
        The above choice of $R_i$ ensures that the expected number of realised connections from $i$ to previous nodes is $m$.
    \end{itemize}
\end{enumerate}
The networks generated according to these rules have the small world property, are scale-free (with a degree decay exponent equal to $1+1/\beta$), and with appropriate choice of $T$ can be made also highly clustered (lower temperature results in larger average clustering coefficient)\cite{PSO}. 
% Kéne még hivatkozni a releváns cikkeket. 
However, an important criticism raised against the PSO model is that for subgraphs spanning between nodes having a degree $k>k_{\rm min}$, we cannot observe the densification law seen in a couple of real networks when $k_{\rm min}$ is increased \cite{Boguna_embedding_2019}. 
%Itt is kéne még hivatkozni.

A generalisation of the PSO model circumventing this problem was proposed in Refs.\cite{PSO,HyperMap}, where the iteration rules listed above are extended by adding extra links also in between already existing nodes as follows:
\begin{enumerate}
    \item[4.] For a randomly chosen, non-connected node pair $j,l<i$ draw a link with probability $p(x_{jl})$, where the hyperbolic distance $x_{jl}$ is calculated from the coordinates $(r_{ji},\theta_j)$ and $(r_{li},\theta_l)$, and $p(x_{jl})$ is evaluated according to equation (\ref{eq:PSO_link_prob}). Repeat this until $L$ number of extra links are created. 
\end{enumerate}

%generalized PSO már a PSO supplementary-jában bevezetve (Section VIII), a HyperMap-ban az újítás az ezzel ekvivalens E-PSO kidolgozása

The main effects of these so-called internal links are that the average degree of the generated network is modified to $\left< k\right>=2(m + L)$, and the average internal degree of the subgraphs between nodes with degrees larger than a certain $k_{\rm min}$ becomes increasing as a function of $k_{\rm min}$ (as discussed in more details in the Supporting Information). The expected total number of internal links from previous nodes on the node appearing at iteration $i$ at the end of the network generation process (assuming altogether $N$ nodes) can be given as \cite{HyperMap}
\begin{equation}
    \bar{L}_i \simeq \frac{2L(1-\beta)}{(1-N^{-(1-\beta)})^2(2\beta -1)}\left[\left(\frac{N}{i}\right)^{2\beta-1}-1\right]\left(1-i^{-(1-\beta)}\right).
    \label{eq:gen_PSO_Li}
\end{equation}

An equivalent model with only external links (connections emerging always with the newly appearing node) was also formulated in Ref.\cite{HyperMap}, which is referred to as the E-PSO model. In this approach we return to the iteration rules 1.--3. of the original PSO model, and omit rule 4. from the generalised version. However, a very important difference compared to the original settings is that the expected number of links connected to the newly appearing nodes is no longer constant, instead it changes during the iterations. In order to obtain on average the same number of links connected to any given node as in the generalised PSO with the internal links, the parameter $m$ in step 3. is replaced by 
\begin{equation}
    m_i = m + \bar{L}_i,
\end{equation}
where $\bar{L}_i$ can be calculated according to equation (\ref{eq:gen_PSO_Li}).

Assuming a network obtained from the E-PSO model, the probability for observing a connection between nodes having the final coordinates $(r_{iN},\theta_i)$ and $(r_{jN},\theta_j)$ at the end of the network generation process was also given in Ref.\cite{HyperMap} in the form of
\begin{equation}
    \tilde{p}(x_{ij}) = \frac{1}{N-i_{\rm min}+1}\sum_{i=i_{\rm min}}^N\frac{1}{1+\mathrm{e}^{\frac{\zeta}{2T}(x_{ij}-R_N+\Delta_i)}}\simeq
    \frac{1}{1+\mathrm{e}^{\frac{\zeta}{2T}(x_{ij}-R_N)}}, 
    \label{eq:global_link_prob}
\end{equation}
where $x_{ij}$ stands for the hyperbolic distance calculated based on equation (\ref{eq:hyper_distance}), $i_{\rm min}=\max(2,\lceil N\mathrm{e}^{-\frac{\zeta x_{ij}}{4(1-\beta)}}\rceil)$, $R_N$ is given by equation (\ref{eq:cutoff}), and $\Delta_i=\frac{2}{\zeta}\ln\left[\left(\frac{N}{i}\right)^{2\beta-1}\frac{mI_i}{m_iI_N}\right]$ with $I_i=\frac{1}{1-\beta}(1-i^{-(1-\beta)})$. Using equation (\ref{eq:global_link_prob}), the likelihood of observing an adjacency matrix $A_{ij}$ for given final hyperbolic distances $x_{ij}$ can be calculated from
\begin{equation}
    \pazocal{L}_{A}  \equiv \pazocal{L}(A_{ij}\mid \{ r_{iN},\theta_i \}, m, L,\zeta,\beta,T)=\prod_{1\leq j <i\leq N} \tilde{p}(x_{ij})^{A_{ij}}\left[1-\tilde{p}(x_{ij})\right]^{1-A_{ij}}.
    \label{eq:likely_A}
\end{equation}
However, when we are interested in the goodness of the fit for an embedding, we need the conditional probability of the node coordinates given the adjacency matrix and the model parameters, which according to Bayes' rule can be expressed as
\begin{equation}
    \pazocal{L}_{r,\theta}\equiv \pazocal{L}_{r,\theta}(\{ r_{iN},\theta_i \} \mid A_{ij} , m, L,\zeta,\beta,T)= \frac{\pazocal{L}(\{ r_{iN},\theta_i \} \mid m, L,\zeta,\beta,T)\cdot \pazocal{L}_{A}}{\pazocal{L}(A_{ij}\mid  m, L,\zeta,\beta,T)},
\end{equation}
where $\pazocal{L}(\{ r_{iN},\theta_i \} \mid m, L,\zeta,\beta,T)$ corresponds to the conditional probability for obtaining the final node coordinates $\{ r_{iN},\theta_i \}$ given the model parameters, and $\pazocal{L}(A_{ij}\mid  m, L,\zeta,\beta,T)$ is the conditional probability for receiving the adjacency matrix $A_{ij}$ given the model parameters. Since the angular coordinates are uniformly random and the radial coordinates (according to the iteration rules 1.--2.) depend only on $\zeta$ and $\beta$, it can be shown that\cite{HyperMap} 
\begin{equation}
    \pazocal{L}(\{ r_{iN},\theta_i \} \mid m, L,\zeta,\beta,T)=
    \pazocal{L}(\{ r_{iN},\theta_i \} \mid \zeta,\beta) = \frac{1}{(2\pi)^N}\prod_{i=1}^N\frac{\zeta}{2\beta}\mathrm{e}^{\frac{\zeta}{2\beta}(r_{iN}-r_{NN})},
\end{equation}
where $r_{NN}=\frac{2}{\zeta}\ln N$. 

If we are given an input network together with model parameters, the maximum likelihood estimate for the node coordinates is formally that set $\{ r_{iN}^*,\theta_i^* \}$ for which $\pazocal{L}_{r,\theta}$ is maximal. As usual, technically it is far more convenient to maximise the logarithm of $\pazocal{L}_{r,\theta}$, which is equivalent to minimising $-\ln \pazocal{L}_{r,\theta}$ given by
\begin{equation}
    -\ln \pazocal{L}_{r,\theta} =
    C - \frac{\zeta}{2\beta}\sum_{i=1}^N r_{iN}-\sum_{i=1}^{N-1}\sum_{j=i+1}^N A_{ij}\ln \tilde{p}(x_{ij})-\sum_{i=1}^{N-1}\sum_{j=i+1}^N(1-A_{ij})\ln\left[1- \tilde{p}(x_{ij})\right],
    \label{eq:negLogLikelihood_def}
\end{equation}
where $C$ is a constant independent from $\{ r_{iN},\theta_i \}$. %and $LL\left(\{ r_{iN},\theta_i \}\right)$ is usually referred to as the logarithmic loss. 
The analytic solution for the optimal radial coordinates can be given as \cite{HyperMap}
\begin{subequations}
\begin{align}
   r_{ii}^* &=\frac{2}{\zeta}\ln i^*, \label{eq:opt_fading_radial} \\
    r_{iN}^* &= \beta r_{ii}^* +(1-\beta)r_{NN}^*, 
    \label{eq:opt_radial}
\end{align}
\end{subequations}
where the optimal ordering of the nodes given by $i^*$ is following the node degrees, with the largest degree node in the network obtaining $i^*=1$, second largest degree node receiving $i^*=2$, etc., and equation (\ref{eq:opt_fading_radial}) corresponds to the initial radial coordinate of node $i^*$, whereas equation (\ref{eq:opt_radial}) takes into account also the outward drift due to the popularity fading. The optimal solution for the angular coordinates cannot be expressed analytically in closed form, opening up the room for heuristic optimisation algorithms. After substituting in equation (\ref{eq:negLogLikelihood_def}) the sum of the $r_{iN}^*$ values expressed from equations (\ref{eq:opt_fading_radial}-\ref{eq:opt_radial}) as a function of the model parameters $\zeta$, $N$ and $\beta$, 
the node arrangement dependent part of the negative log-likelihood can be written as
%that part of the negative log-likelihood which still remains dependent on the actual node arrangement can be written as
%Looking back at equation (\ref{eq:negLogLikelihood_def}), we can see that the coordinate dependent part of the log-likelihood can be written as
\begin{equation}
    LL \equiv -\ln \pazocal{L}_{A} = -\sum_{i=1}^{N-1}\sum_{j=i+1}^N A_{ij}\ln \tilde{p}(x_{ij})-\sum_{i=1}^{N-1}\sum_{j=i+1}^N(1-A_{ij})\ln\left[1- \tilde{p}(x_{ij})\right],
    \label{eq:logloss_def}
\end{equation}
which we shall refer to as the logarithmic loss from here on.

Probably the most well-known method for minimising the logarithmic loss is HyperMap, introduced in Ref.\cite{HyperMap} for embedding networks based on the E-PSO model. In this approach the nodes of the network are sorted and indexed in decreasing order of their degree. The node with the largest degree (indexed by $i=1$) is placed at the centre of the hyperbolic disk, and the rest of the nodes are introduced one by one, obtaining initial radial coordinates given by equation (\ref{eq:opt_fading_radial}). At the introduction of a new node, the radial coordinates of the previous nodes are updated according to the concept of popularity fading, and the angular coordinate of the new node is chosen by minimising a local version of the logarithmic loss, where contributions only from the already introduced nodes (including the new node) are taken into account. Further details of the algorithm are given in Ref.\cite{HyperMap}.

\subsection{Coalescent embedding with ncMCE}

The short outline of the coalescent embedding methods is the following: first a weighted adjacency matrix is prepared (this step can be referred to as pre-weighting), based on which the node similarity matrix $\mathbf{D}$ is obtained, and then the angular coordinates of the nodes are gained by applying a dimension reduction technique to $\mathbf{D}$ \cite{coalescentEmbedding}. The rationale behind this approach is that when applied to a network that is known to be hyperbolic, a common node aggregation pattern can be observed in the embedding space which is circularly or linearly ordered (angular coalescence) according to the original angular coordinates in the hyperbolic space. An extensive study of different similarity matrices and dimension reduction methods was carried out in Ref.\cite{coalescentEmbedding}, and according to tests on real input networks, the best greedy routing scores could be achieved by combining repulsion-attraction (RA) pre-weighting with ncMCE dimension reduction.

In this approach we first prepare a weighted adjacency matrix $\mathbf{W}$ with elements
\begin{equation}
    W_{ij} = \frac{k_i+k_j+k_ik_j}{1+CN_{ij}},
    \label{eq:RA_W}
\end{equation}
where $k_i$ and $k_j$ denote the degree of nodes $i$ and $j$, and $CN_{ij}$ stands for the number of common neighbors of these two nodes. The appearance of $CN_{ij}$ in the denominator of equation (\ref{eq:RA_W}) provides a sort of 'repulsion' between nodes having neighbours not in common (large $k_i$ and $k_j$ compared to $CN_{ij}$), resulting in larger $W_{ij}$ values, which reflect less similarity. Next the minimum weight spanning tree of the induced weighted network is prepared, and the entries of the similarity matrix $\mathbf{D}$ are given by the distance of the corresponding node pair in the spanning tree. The matrix element $D_{ij}$ can be interpreted as an estimate for the minimum curvilinear distance between node $i$ and node $j$ \cite{ncMCE,ncMCE_early}.

The dimension reduction is carried out via singular value decomposition, corresponding to a factorisation of $\mathbf{D}$  as $\mathbf{D}=\mathbf{U}\bm{\Sigma}\mathbf{V}^T$, where $\bm{\Sigma}$ is a diagonal matrix containing the singular values, from which we keep only the two largest ones (and put the rest to zero) in the following. Given a distance matrix, multidimensional scaling (MDS) can be used to determine that set of node coordinates which gives back all the pairwise distances. In our case the angular coordinates of the nodes are obtained from the matrix $\mathbf{X} = \left(\sqrt{\bm{\Sigma}}\cdot\mathbf{V}^T\right)^T$. Although already the coordinates in the second column of $\mathbf{X}$ could be regarded as the angular coordinates when re-scaled into the interval $[0,2\pi)$, according to Ref.\cite{coalescentEmbedding} we can obtain better results by applying an equidistant adjustment. Technically this is equivalent to distributing the angular coordinates in a regular uniform fashion over the interval $[0,2\pi)$, following the node order dictated by the second column of $\mathbf{X}$. The radial coordinates are obtained in the same way as in the logarithmic loss optimising methods, making use of equations (\ref{eq:opt_fading_radial}-\ref{eq:opt_radial}), where the radial order of the nodes is adjusted according to their degree.

\subsection{Optimisation of the angular coordinates}

As mentioned in the Introduction, our embedding method combines the coalescent embedding approach with an optimisation of the angular coordinates based on the assumption that the network to be embedded was generated according to the E-PSO model. In the first state of the embedding process we apply the RA pre-weighting given by equation (\ref{eq:RA_W}) for preparing the similarity matrix $\mathbf{D}$, and use the ncMCE dimension reduction technique described in the previous section to obtain the coordinate matrix $\mathbf{X}$. The initial angular coordinates inputted to our optimising algorithm correspond to the elements in the second column of $\mathbf{X}$ after equidistant adjustment. 

During the optimisation we iterate over the network nodes according to their radial order (beginning with the innermost node), and examine in each iteration a $q=6$ number of new angular positions for the current node, which are placed equidistantly between the second neighbours of the node according to the (current) angular node order. For each examined new position the logarithmic loss given by equation (\ref{eq:logloss_def}) is calculated, and if lower values are observed compared to the original one, the angular coordinate of the current node is updated to the best new position. The reason for limiting the arc of possible new positions between the two second neighbours is that the original coordinates obtained with the ncMCE method are usually already quite good; thus, only minor adjustments are needed for improving the embedding. Nevertheless, with this choice of boundaries we also allow swaps in the angular order of the nodes. (Whenever we have to update the angular coordinate of the current node to a new position in between the first and second neighbours, the angular order is changed).

Let us denote one iteration over all nodes as described above as a swapping round (due to the possibility of swaps in the angular order). After a few of these swapping rounds, in order to enable the settling of the angular positions to the true optimum allowed by the current angular order, we carry out a couple of non-swapping rounds of updates, where the $q$ number of possible new angular positions are distributed only between the first angular neighbours of the current node. (E.g., in our experiments on synthetic networks, we used 5 swapping rounds followed by 3 non-swapping rounds.) The total number of rounds $n$ can be either preset, or applying a stop condition based on the relative improvement in $LL$ over the consecutive rounds is also a simple option.

In terms of complexity, the calculation of the change in the logarithmic loss $LL$ when trying out a new node position involves the evaluation of $N-1$ number of terms, consequently the total number of calculation steps needed to perform the angular optimisation is proportional to $n\cdot N\cdot q\cdot(N-1)$. This means that the running time of our algorithm is linear in terms of $n$ and $q$, and quadratic in terms of $N$. Thus, keeping low the number of optimisation rounds and the number of test positions per node compared to the network size $N$, the computational complexity of the proposed embedding optimisation method is $\mathcal{O}(N^2)$, similarly to that of the original coalescent embedding approach based on ncMCE dimension reduction \cite{coalescentEmbedding}. 

Before actually showing the results of our algorithm, it is important to specify how we choose the parameters $\zeta,\,m,\,L,\,\beta$ and $T$ of the logarithmic loss, that are considered to be fixed during the angular optimisation. Following the standard practice in the literature, $\zeta$ (characterising the curvature of the hyperbolic plane) is always assumed to be $1$. For $m,L,\beta$ and $T$ we already mentioned in the description of the PSO and E-PSO models that these parameters are connected to the different statistical features of the generated graphs in mostly simple forms (e.g., the average degree is $\left< k\right>=2(m+L)$, etc.); thus, a reasonable estimate for these can be made by observing the corresponding properties of the network to be embedded. However, in order to obtain the best results, in our studies we actually optimise for these parameters as well, prior to the optimisation of the angular coordinates. 

As a first step,  we apply the ncMCE based coalescent embedding to obtain some initial coordinates for the nodes. Using these, we optimise $m,\,\beta$ and $T$ simultaneously by minimising the logarithmic loss in equation (\ref{eq:logloss_def}) via a simple gradient descent in the $m-\beta-T$ parameter space (while the node coordinates are kept fixed). The value of $L$ is calculated then from the relation $\left< k\right>=2(m+L)$; hence, it is actually not regarded as a free parameter on its own. However, an important feature of the parameter optimisation is that we allow $L$ to take negative values as well, or in other words, $m$ is allowed to take values larger than $\left< k\right>/2$. This corresponds to a further generalisation of the E-PSO approach, where during the graph generation process in step 4. we actually delete already existing  links (with probabilities depending on the hyperbolic distance) instead of adding extra links. A more detailed discussion of the generalised $L$ parameter and the tuning of the embedding parameters is provided in the Supporting Information.

We have two important final remarks related to the parameters of the embedding, concerning the radial ordering of the nodes dictated by the node degrees. First, in the case of directed networks we can associate 3 (in principle different) degree values to each node, namely the in-, the out- and the total degree (corresponding to the sum of the in- and the out-degree). Thus, for any directed network we can construct at least 3 different radial orderings depending on the degree type we use when ranking the nodes. According to our knowledge, no systematic study has been carried out to investigate which degree type should be preferred over the others in general; therefore, here we take a practical approach by trying out all 3 possibilities whenever dealing with directed networks and choose the ordering that yields the best quality scores. 

Our second remark is related to the very likely ambiguity in the radial ordering for any network (both directed and un-directed) caused by the occurrence of equal node degrees in the system. I.e., in real networks the degree distribution is usually skewed, meaning that a relatively large fraction of the nodes has a small degree compared to the average degree. This means that in the low degree regime we usually find a considerable number of nodes with the very same degree, hence the radial ordering dictated by equations (\ref{eq:opt_fading_radial}-\ref{eq:opt_radial}) allows actually a large number of different permutations within segments of the node ranking containing nodes with equal degree. According to our experiments detailed in the Supporting Information, there can be a non-negligible variance in the quality scores measuring the goodness of the embedding when permuting the radial order between nodes of the same degree for both HyperMap, the original ncMCE approach, and also the optimised ncMCE method proposed in this paper. Therefore, the actually chosen radial order (out of the many possibilities that are monotonic according to the degree) can be viewed as a further parameter of the embedding for the aforementioned methods. However, the optimal choice for this parameter can be set only via trial and error, i.e. by repeatedly trying out different random permutations between the nodes of the same degree, and keeping that radial order which produces the best quality score. Under some circumstances an estimate on the possible further improvement in the quality score as a function of the number of further tries can be made, as shown in the Supporting Information.

\section{Results}
We have tested our method on both synthetic and real networks (our code is available from Ref.\cite{our_code}). In order to quantify the quality of the embedding, we used the logarithmic loss defined in equation (\ref{eq:logloss_def}), and also the greedy routing score, which is a commonly used, model-free measure \cite{HyperMap,coalescentEmbedding}. The idea of greedy routing on a network embedded in a geometric space corresponds to a simple routing protocol for getting from a source node $i$ to a destination node $j$ by walking on the network, where the next step from the current node is always carried out to the neighbour that is the closest to the destination $j$ according to the distance measured in the given geometric space \cite{Kleinberg_greedy_routing}. For networks embedded in a hyperbolic space, the distance we use during greedy routing is the hyperbolic distance between the nodes. If a node has no neighbours closer to the destination compared to itself, the path is stopped. Therefore, a natural simple measure for the success of the routing protocol is given by the fraction of successful paths actually reaching the destination without getting stuck on any other node \cite{Boguna_2009_nat_phys}. In order to measure the success of the routing in a more refined way, we calculate the greedy routing score \cite{coalescentEmbedding} given by
\begin{equation}
    GR = \frac{1}{N(N-1)}\sum\limits_{i=1}^N\sum\limits_{j=1,j\neq i}^N \frac{\ell_{ij}^{(SP)}}{\ell_{ij}^{(GR)}},
\end{equation}
where $\ell_{ij}^{(SP)}$ denotes the shortest path length between $i$ and $j$, and $\ell_{ij}^{(GR)}$ stands for the greedy routing path length between the same source-destination pair, which is considered to be infinity if the routing fails in reaching $j$ from $i$.

\begin{figure}[hbt]
    \centering
    \includegraphics[width=\textwidth]{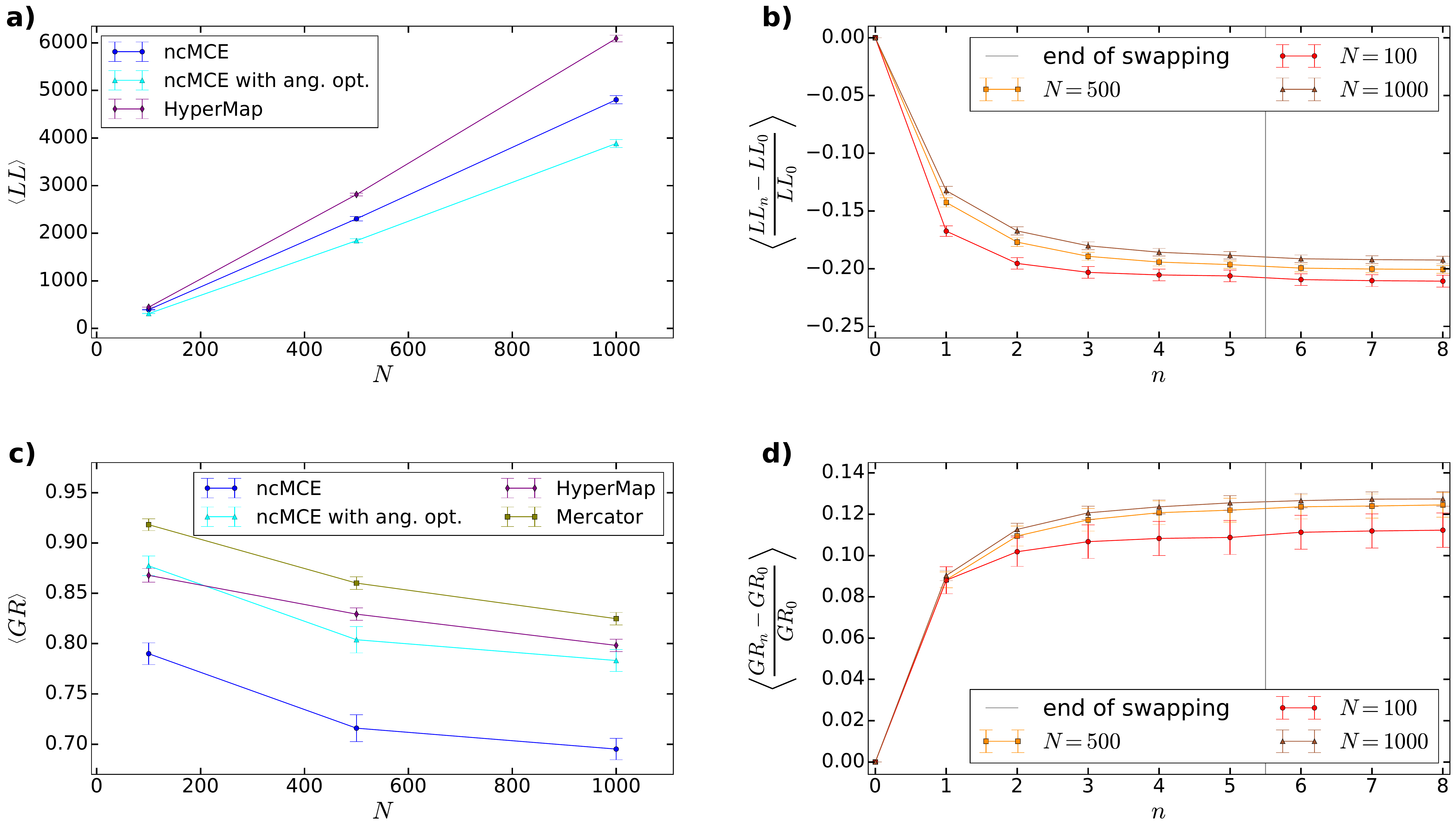}
    \caption{ {\bf Embedding results for synthetic networks.} a) The average logarithmic loss $\left< LL\right>$ and the corresponding 95\% confidence interval (indicated by bars) as a function of the number of nodes $N$ for 100 networks generated by the PSO model using $\zeta=1$, $m=2,\beta=2/3$ and $T=0.3$. b) The convergence of the logarithmic loss over the subsequent rounds of iterations during the proposed angular optimisation of the ncMCE method. c) The average greedy routing score $\left< GR\right>$ and the corresponding 95\% confidence interval (indicated by bars) as a function of the number of nodes $N$ for the same synthetic data set as in panel a). d) The convergence of the greedy routing score as a function of the number of rounds $n$ during our angular optimisation.}
    \label{fig:synthetic_logloss}
\end{figure}

In Fig. \ref{fig:synthetic_logloss} we show the results for synthetic networks generated by the PSO model with sizes $N=100,\,500$ and $1000$. We tested four embedding methods on 100 networks at each network size. In Fig. \ref{fig:synthetic_logloss}a we plot the average logarithmic loss $\left< LL\right>$ as a function of $N$ for HyperMap (purple), the original ncMCE (blue) and ncMCE with angular optimisation (cyan). (Since Mercator is based on the $\mathbb{S}^1/\mathbb{H}^2$ model, the logarithmic loss with respect to the E-PSO model cannot be considered as a fair quality function regarding this embedding method; therefore, Mercator is left out from Fig. \ref{fig:synthetic_logloss}a.) Not surprisingly, the curves show an increasing tendency with $N$; however, the angular optimisation clearly provides an about $20\%$ lower $LL$ compared to ncMCE without optimisation, and about a $30\%$ lower value compared to HyperMap. 
In Fig. \ref{fig:synthetic_logloss}b we show the relative change in $LL$ as a function of the number of rounds $n$ during our optimisation of the angular coordinates. According to the figure, the $LL$ seems to settle to a more or less constant value after $6-8$ rounds. In Fig. \ref{fig:synthetic_logloss}c we display the average greedy routing score $\left< GR\right>$ as a function of the system size $N$. This figure indicates that the angular optimisation improves the result of ncMCE in terms of $GR$ as well; however, the greedy routing score obtained with HyperMap is not surpassed, and the best greedy routing scores are obtained with Mercator. In Fig. \ref{fig:synthetic_logloss}d we plot the relative change in $GR$ as a function of the number of rounds $n$ in the angular optimisation of the result of ncMCE, where -- similarly to Fig. \ref{fig:synthetic_logloss}b -- a steady value is reached roughly above $n=6$. Additional figures related to embedding results on synthetic networks are provided in the Supporting Information.

In terms of real systems, we tested our method on the Pierre Auger collaboration network ($N=38$ nodes, available from Ref.\cite{Pierre_Auger_data}), a network between books about US politics, where links correspond to frequent co-purchasing ($N=105$ nodes, available from Ref.\cite{pol_books_data}), the American College Football network ($N=115$, available from Ref.\cite{football_net_data}), a Cambrian food web from the Burgess Shale ($N=142$, available from Ref.\cite{foodweb}) and a protein interaction network from the PDZBase database ($N=161$, available from Ref.\cite{pdz_base_net}). An important note about the Cambrian food web is that this network is usually considered to be directed, where links are pointing from consumers to their food resources. According to that, we tried out all 3 options for defining the radial order among the nodes as described in the previous section for both HyperMap, the original ncMCE based coalescent embedding and our algorithm (whereas Mercator does not allow this option). The comparison between the 3 options showed non-trivial results, e.g. the ordering according to the in-degree achieved the best greedy routing scores for both the original ncMCE approach and our algorithm, although its score was surpassed by the ordering according to the total degree in the case of HyperMap. More details on this aspect of the Cambrian food web are given in the Supporting Information.
%According to that, each node can be characterised by its in-degree, out-degree, and total degree (corresponding to the sum of the previous two quantities), raising the question which of these should be used for determining the radial order between the nodes? In our studies we tried out all 3 options for both HyperMap, the original ncMCE based coalescent embedding and our algorithm (whereas Mercator does not allow this option). 

\begin{figure}
    \centering
    \includegraphics[width=\textwidth]{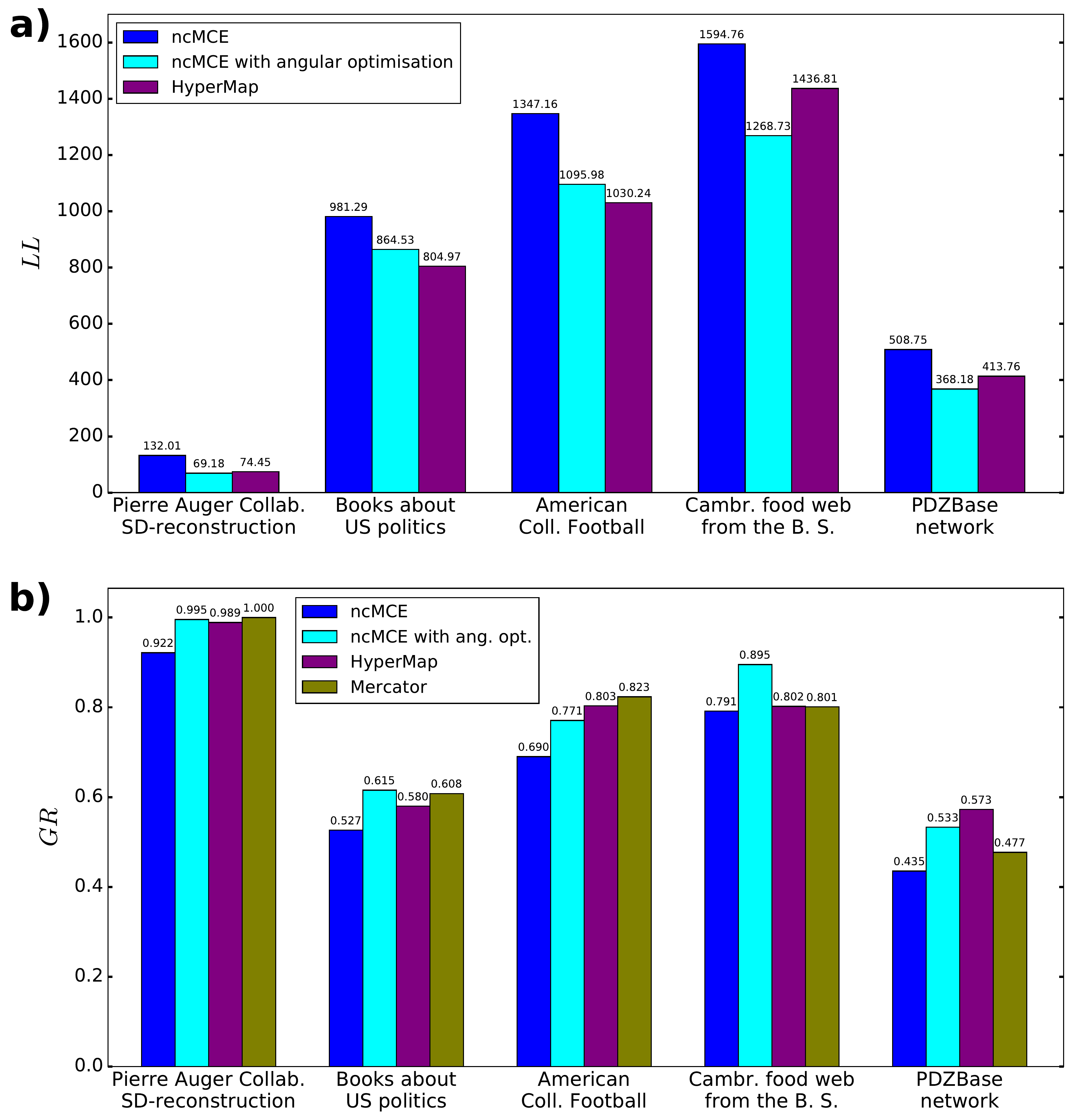}
    \caption{{\bf Embedding results for real networks.} a) The logarithmic loss for HyperMap (purple), the original ncMCE (blue) and ncMCE with angular optimisation (cyan) for the real networks that we studied. b) The greedy routing score for the same methods and also Mercator (olive).}
    \label{fig:real_barplots}
\end{figure}

In Fig. \ref{fig:real_barplots} we show a summary of the quality scores obtained for the real networks, displaying the best results we could achieve for each method depending on the choice of the embedding parameters. For each network we performed embeddings trying out $2500$ radial orders of the nodes with each $m-L-\beta-T$ parameter setting for HyperMap, the original ncMCE and our approach, and we also embedded each network $2500$ times with Mercator (where the repeated embedding of the same network also provides varying results). The total number of rounds $n$ needed in our optimisation framework varied between $n=8$ and $n=20$ for the studied real networks (details are given in the Supporting Information).
%lehet állítani paramétereket, csak nem azokat, mint a többi módszernél, pl. a radiális sorrendet nem (mert itt nincs is olyan)
%(where the parameters cannot be adjusted, but nevertheless, repeated embedding of the same network provides varying results).
In panel a) we compare the logarithmic loss $LL$ in our approach to the results of HyperMap and the original ncMCE coalescent embedding, and in panel b) we show the greedy routing score $GR$ for the same methods including the results of Mercator as well. (Mercator is left out from panel a) for the same reason as in the case of Fig. \ref{fig:synthetic_logloss}). %Naturally, the value of the quality scores can depend on the parameters of the given method; here we show the best results we obtained for each involved method over different choices for the parameters. 
Further results at different parameter settings (together with a more detailed description of the studied networks) are given in the Supporting Information.

According to Fig. \ref{fig:real_barplots}a, our angular optimisation reduces the logarithmic loss compared to the value obtained with the original ncMCE approach also for real networks. The maximum reduction was $47.6\%$ (for the Pierre Auger collaboration network) and the average reduction was $25.24\%$ for the studied systems. Our method obtained the lowest $LL$ in $3$ out of the $5$ cases, and the aforementioned reduction played a quite important role in this, since the logarithmic loss of the original ncMCE turned out to be higher compared to that of HyperMap. Besides, the proposed optimisation increases the greedy routing score, as shown in Fig. \ref{fig:real_barplots}b. The maximum improvement compared to the original ncMCE in terms of $GR$ was $22.5\%$ (for the protein interaction network from the PDZBase database), and the average improvement was $14.41\%$ for the studied systems. In addition, our approach achieved the highest $GR$ for $2$ out of the $5$ studied networks, and the second best greedy routing score in $2$ more cases.

\begin{figure}
    \centering
    \includegraphics[width=\textwidth]{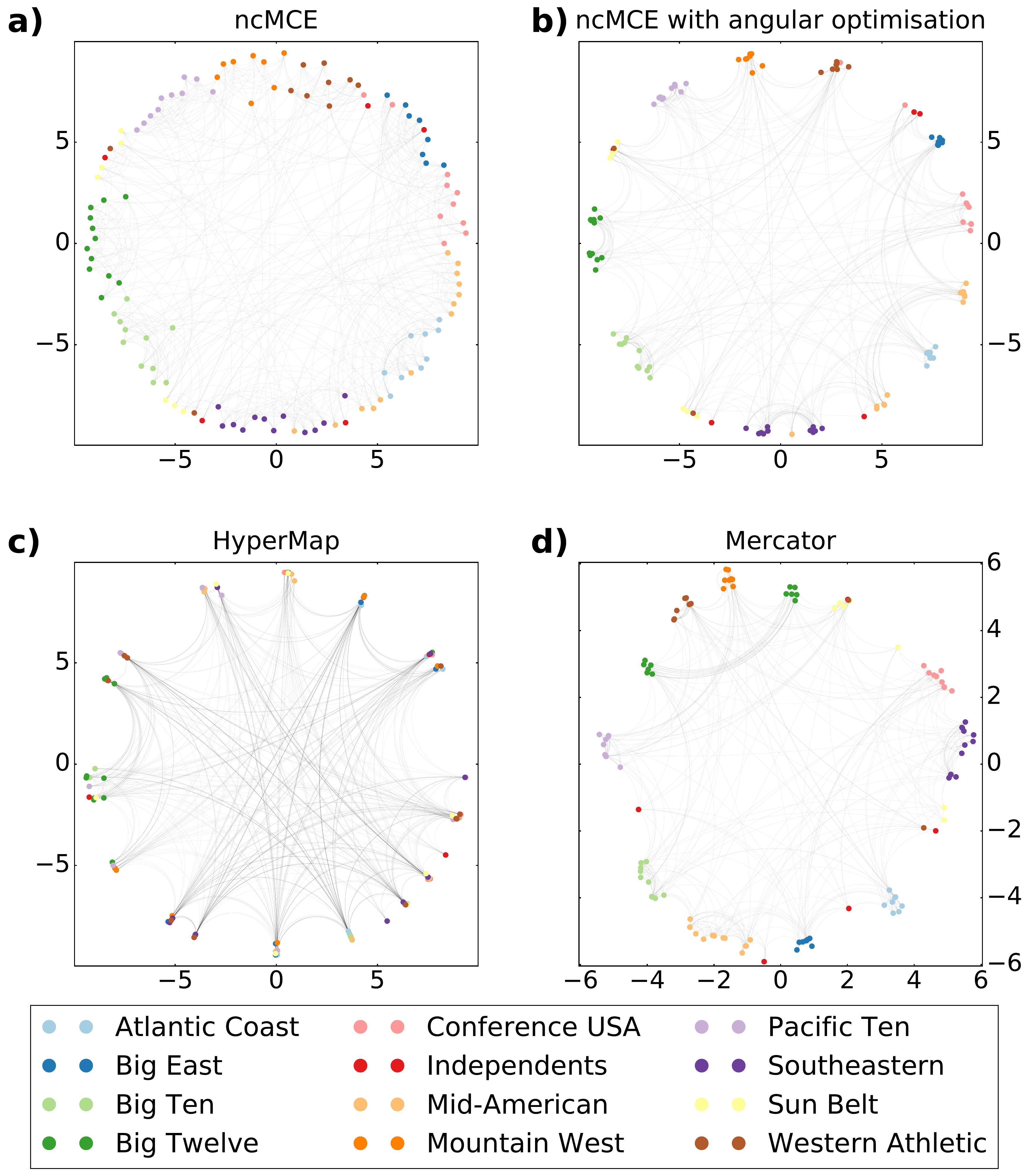}
    \caption{ {\bf The layouts of the American College Football web on the native hyperbolic disk that reached the highest greedy routing scores.} a) The layout based on the coordinates resulted from the original ncMCE method. b) The layout according to the coordinates obtained with our approach, optimising the results of ncMCE. c) The hyperbolic layout obtained with HyperMap. d) The embedding according to Mercator.}
    \label{fig:PoincDisk}
\end{figure}

In Fig. \ref{fig:PoincDisk} we compare the layouts of the American College Football web in the 2D hyperbolic space obtained with the four different embedding methods. An interesting feature of this data set is that information about the conferences of the included teams is also available, which is marked by the different node colours in the figure. The angular coordinates of the nodes are equidistantly distributed in the output of the original ncMCE based coalescent embedding approach, as it can be seen in Fig. \ref{fig:PoincDisk}a. A visually quite pleasant feature of this layout is that according to the colouring, the teams belonging to the same conference tend to occupy a more or less well-defined, continuous range according to the angle. After applying the angular optimisation proposed in this paper, the angular coordinates are no longer equidistantly distributed and -- as shown in Fig. \ref{fig:PoincDisk}b -- the conferences contract into well-separated clusters, which helps the viewer even more in separating the different groups during a visual interpretation of the layout.
HyperMap seems to repeatedly assign the same (or very close) angular coordinates for multiple nodes at the same time, which results in very tight clusters in the layout (Fig. \ref{fig:PoincDisk}c); however, according to the colouring of the nodes, these clusters often contain nodes from different conferences. The layout obtained with Mercator (Fig. \ref{fig:PoincDisk}d) shows an organisation similar to that of our algorithm, where most of the team conferences appear as well-separated, but not extremely tight clusters. 

\begin{figure}
    \centering
    \includegraphics[width=0.9\textwidth]{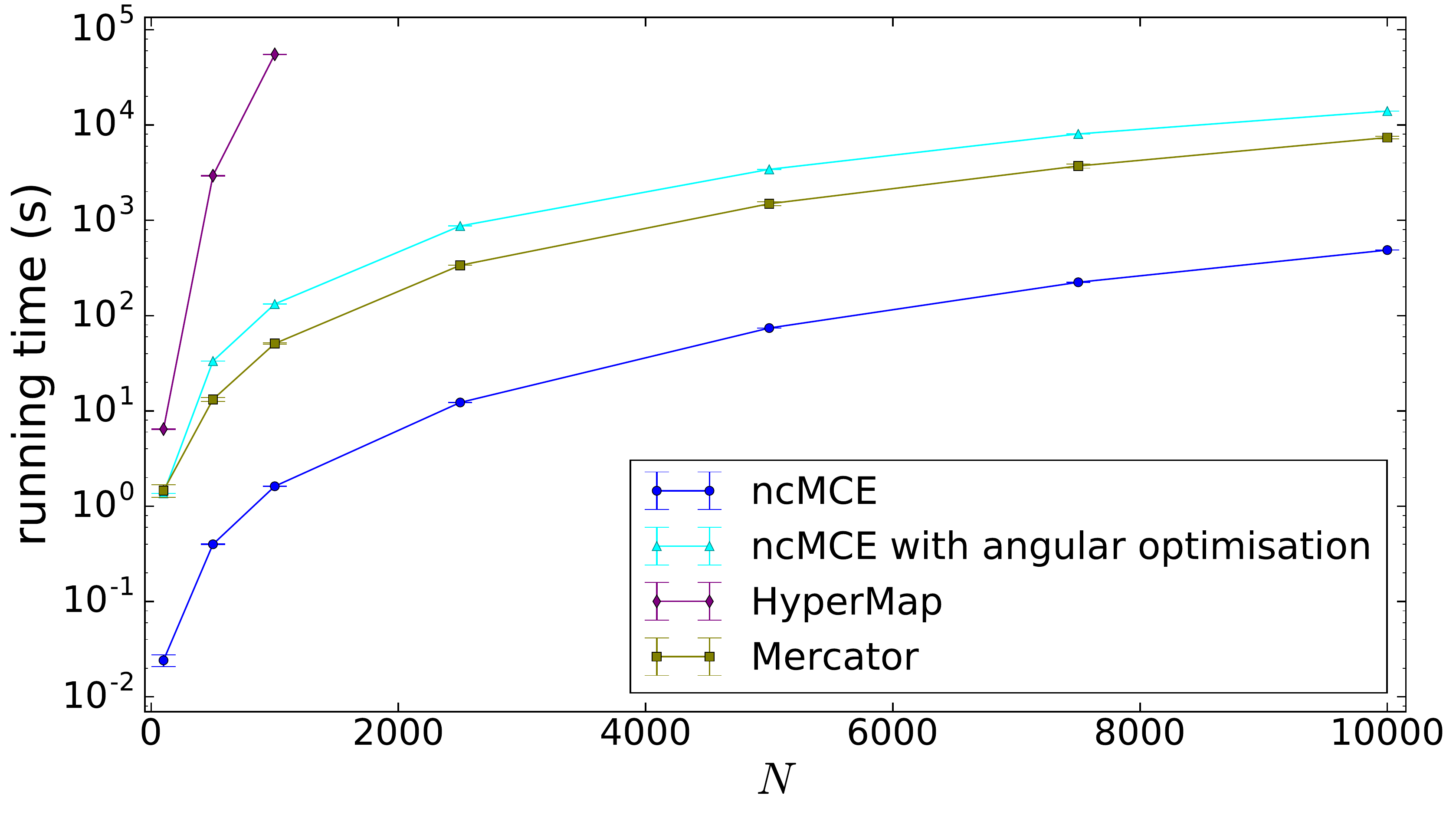}
    \caption{{\bf The running times of the studied embedding algorithms.} The tests were run on synthetic networks obtained from the PSO model with $\zeta=1,\,m=2,\,\beta =2/3,\,T=0.3$ and $N$ increasing from $100$ to $10,000$, where $10$ samples were generated at each $N$ value. The measured average running time is shown in seconds, the bars indicate the 95\% confidence intervals.}
    \label{fig:running_times}
\end{figure}

In Fig. \ref{fig:running_times} we show the running time for the different embedding algorithms, measured for synthetic networks generated by the PSO model. The size of these networks varied between $N=100$ and $N=10,000$ nodes, while the further parameters of the network generating model were kept fixed.
The original ncMCE algorithm seems to be the fastest, with Mercator coming second followed quite closely by our approach, and HyperMap appears to be way slower compared to the rest at larger network sizes.

\section{Discussion}
The coalescent hyperbolic embedding based on the ncMCE dimension reduction was shown to be a very efficient method with low running time and high quality results \cite{coalescentEmbedding}. In the present work we proposed a further optimisation of the angular coordinates obtained with this approach using a logarithmic loss function based on the E-PSO model. According to our experiments on both synthetic and real networks, this comes with the cost of a somewhat increased running time, but it also provides a lower logarithmic loss $LL$ and a higher greedy routing score $GR$. The reduction of $LL$ is not at all surprising (since we are actually optimising with regard to that); however, the improvement in $GR$ in the meantime is indicating that the embedding becomes better also according to a model-free quality score. In addition to the original ncMCE approach, we compared the results of our algorithm also with embeddings obtained with HyperMap \cite{HyperMap} and Mercator \cite{Boguna_embedding_2019}, and it seems that our algorithm is competing with these state-of-the-art methods in terms of the aforementioned two quality scores.

In the case of the American College Football web the optimisation of the angular coordinates led to a result where clusters of nodes belonging to the same team conferences became more separated from the other groups compared to the layout in the original ncMCE approach. This shows that in some cases our algorithm not only improves the quality score of the embedding, but it can in addition provide a layout that is more intuitive and easy to interpret. Based on the above, the usage of our extension of the ncMCE coalescent embedding can be quite beneficial in any further study or application where high quality hyperbolic embedding of networks is important.

A final remark we would like to make regarding the quality of the embedding (measured by either the logarithmic loss or the greedy routing score) is related to the radial order of the nodes dictated by the node degree in HyperMap, the original ncMCE coalescent embedding and also in our approach. As mentioned previously, real networks are very likely to contain (in some cases even large) groups of nodes with equal degree, and within such a group the radial order of the nodes can be chosen arbitrarily. According to our analysis (detailed in the Supporting Information), depending on the actual choice of the radial order the quality scores can show a non-negligible variance. Furthermore, in directed networks in principle we can choose from $3$ degree types (corresponding to the in-, the out- and the total degree) when defining an ordering among the nodes. Our studies related to the Cambrian food web showed that the quality scores are sensitive also to the chosen degree type defining the radial order. These features open up additional possibilities for optimisation to be studied in further works.

%\section*{Methods}

\bibliographystyle{unsrt}
\bibliography{references}

\section{Acknowledgements}

The research was partially supported by the European Union through projects ‘RED-Alert’ (grant no.: 740688-RED-Alert-H2020- SEC-2016-2017/H2020- SEC-2016-2017-1) and by the Hungarian National Research, Development and Innovation Office (grant no. K 128780, NVKP\_16-1-2016-0004). The funders had no role in study design, data collection and analysis, decision to publish, or preparation of the manuscript.

\newpage

\begin{center}
\LARGE{\bf SUPPORTING INFORMATION}
\end{center}

\renewcommand{\thefigure}{S\arabic{figure}}
\renewcommand{\thetable}{S\arabic{table}}
\renewcommand{\theequation}{S\arabic{equation}}
\renewcommand{\thesection}{S\arabic{section}}

\setcounter{section}{0}
\setcounter{figure}{0}
\setcounter{equation}{0}
\setcounter{table}{0}

\section{The E-PSO model}

The popularity-similarity optimisation (PSO) model\cite{PSO} was introduced in order to describe how random geometric graphs grow in the native disk representation of the hyperbolic plane. Its importance derives from the fact that networks generated by this model exhibit scale-free degree distribution, strong clustering and also the small world property, which are frequently mentioned among the universal properties of real networks. However, as mentioned in the main article, an important criticism raised against the PSO model is that the densification law for subgraphs spanning between nodes having a degree $k>k_{\rm min}$ cannot be observed when $k_{\rm min}$ is increased, as opposed to some of the real networks \cite{Boguna_embedding_2019}. The generalisation offered by the E-PSO model solves this problem, as we show here.

The essential idea realised by the PSO model is that at each time step a new node appears in the hyperbolic plane (with a radial coordinate depending on the time and a random angular coordinate) and it establishes connections with some of the previously appeared nodes, which are selected according to probabilities determined by their hyperbolic distance from the new node. The links obtained this way -- connecting always the new node to old ones -- are called \it external \rm links. However, it is plausible to assume that new connections may emerge during the network growth also between pairs of old nodes. The PSO model can be easily extended with the formation of such \it internal \rm links\cite{PSO}: in the so-called\cite{HyperMap} generalised PSO model in addition to the $m$ number of external links introduced by the new node, at each time step an internal link is also created between $L_+$ number of disconnected node pairs, which are sampled according to distance-dependent probabilities.

The E-PSO model\cite{HyperMap} is an equivalent of the generalised PSO model described above, simulating the emergence of the $L_+$ number of internal links by creating \textit{different number} of external links at each time step. In this approach, for a given total number of nodes $N$ and popularity fading parameter $\beta$, in expected value
\begin{equation}
    \bar{m}_i = m+\bar{L}_i \simeq m+L_+\cdot\frac{2(1-\beta)}{(1-N^{-(1-\beta)})^2(2\beta -1)}\left[\left(\frac{N}{i}\right)^{2\beta-1}-1\right]\left(1-i^{-(1-\beta)}\right)
    \label{eq:gen_PSO_mLi}
\end{equation}
new links are created at time $i$ (i.e. at the appearance of node $i$), each of them connecting a previously appeared node (indexed by $j<i$) to the new node $i$.

To generalise the concept of internal links further, it is also conceivable that after a while some of the connections are deleted. Along this line, we can extend the generalised PSO model with the deletion of the link between $L_-$ number of connected pairs of old nodes at each time step. But how should the links be selected for deletion? If the temperature $T$ is set to $0$, when creating new (either external or internal) links we have to always connect that node pair from the candidates which is characterised by the smallest hyperbolical node-node distance. The opposite of this deterministic connection rule is easy to phrase: for $T=0$ in each deletion step the link connecting the hyperbolically furthermost nodes is split up. Consequently, for $T=0$ the case $L_+=L_-$ gives back exactly the original PSO model.

At $T>0$ a natural extension of the above concept is to assume such a link removal process where the probability that a link will not be deleted corresponds to the usual PSO linking probability, and the complementary probability of this is the removal probability, according to which we remove at each time step $L_-$ number of internal links at random. In this way, when $L_+=L_-$ in the generalised PSO model, we add and remove the same number of internal links at each time step, and therefore, the resulting networks become equivalent to the networks generated by the original PSO model.
%Tényleg ekvivalencia van a PSO-val? Oké, hogy összesen minden lépésben m új élt csinálunk itt is, de nem feltétlenül u.azokat, mint amelyeket egy PSOnál csinálnánk (ugyan valószínűleg a nagy hiperb.táv.-ú éleket töröljük, de csak valószínűleg; bár ugyanígy a PSOnál is csak valószínűleg a rövid éleket húzzuk be 0<T-re)

By taking $L=L_+ - L_-$ as the net number of added and removed internal links per time step, we can also consider the analogous generalised E-PSO model, where all connections are created as external links at the appearance of the new nodes, without any additional link insertion or deletion. In this framework, by adjusting $\bar{m}_i$, the expected number of new links introduced connected to the new node $i$, the resulting network can be made equivalent to the generalised PSO model inserting and deleting internal links. The method is straightforward, we can simply use 
\begin{equation}
    \bar{m}_i = m+\bar{L}_i \simeq m+L\cdot\frac{2(1-\beta)}{(1-N^{-(1-\beta)})^2(2\beta -1)}\left[\left(\frac{N}{i}\right)^{2\beta-1}-1\right]\left(1-i^{-(1-\beta)}\right),
    \label{eq:gen_PSO_gen_Li}
\end{equation}
where the only (but important) difference compared to equation (\ref{eq:gen_PSO_mLi}) is that $L$ can also be negative, whereas $L_+$ in equation (\ref{eq:gen_PSO_Li}) is always non-negative.

\begin{figure}[hbt]
    \centering
    \includegraphics[width=\textwidth]{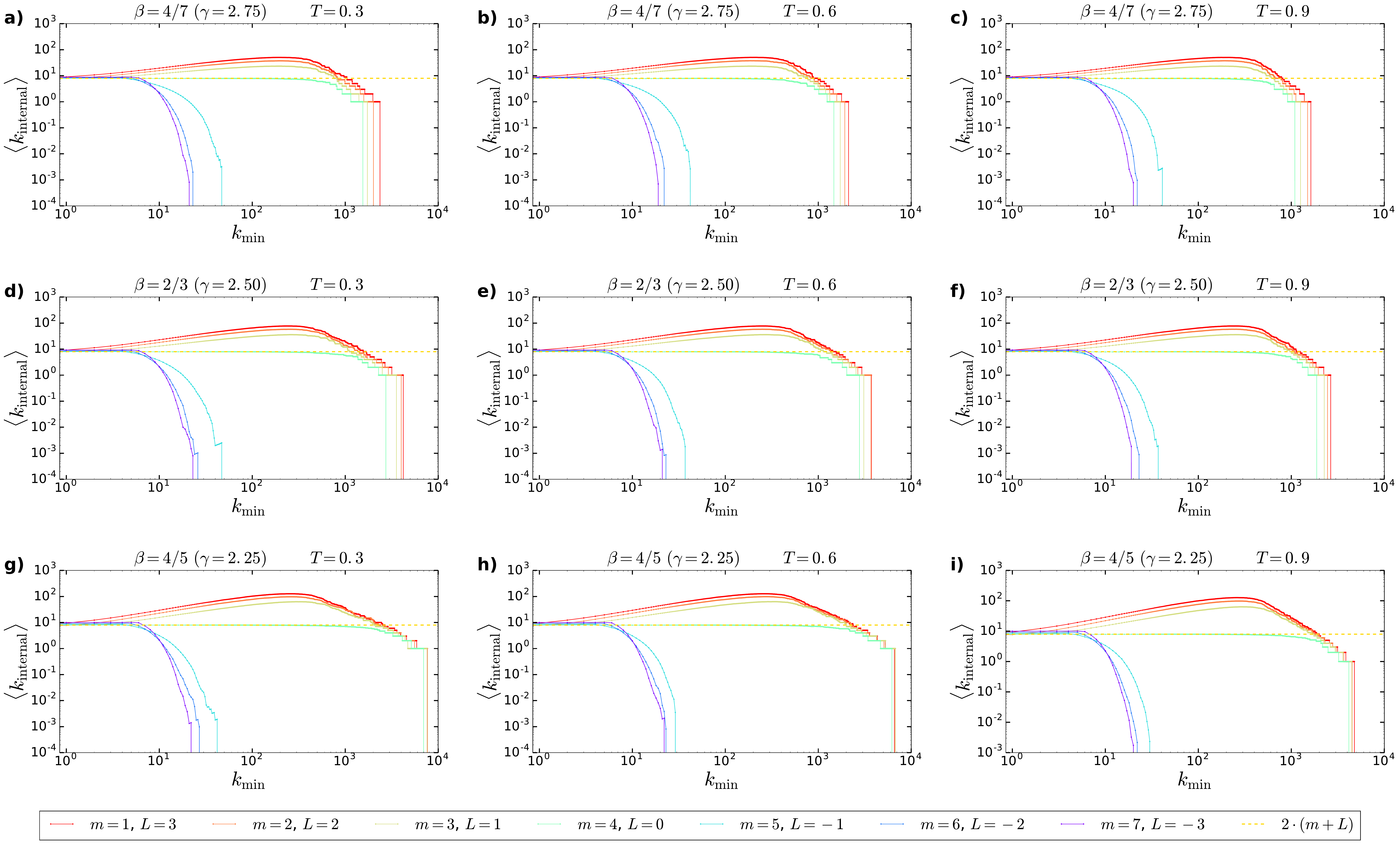}
    \caption{{\bf Average internal degree of subgraphs spanning between nodes with degrees larger than a threshold as a function of the degree threshold for synthetic networks generated by the E-PSO model using different parameters.} With each parameter setting one network was generated with size $N=100000$. The parameter $\zeta$ was always set to $1$. Each panel corresponds to a certain $\beta-T$ setting given in the title of the subplot. The different colours of the curves indicate the different settings of the parameters $m$ and $L$. The expected average degree $2\cdot(m+L)$ was $8$ for each network.}
    \label{fig:EPSOavDegCurve}
\end{figure}

In Fig. \ref{fig:EPSOavDegCurve} we plot the  average internal degree $\left< k_{\mathrm{internal}}\right>$ of the subgraphs spanning between nodes having a degree larger than a certain threshold $k_{\rm min}$ as a function of $k_{\rm min}$ for both positive and negative $L$ values (indicated by different colours) at different $\beta$ and $T$ parameters.
%demonstrate that the shape of the curve describing the dependency of the average internal degree $\left< k_{\mathrm{internal}}\right>$ of the subgraphs spanning between nodes having a degree larger than a certain threshold $k_{\rm min}$ on the value of the degree threshold can be controlled in the E-PSO model by adjusting the parameter $L$.
When $L$ is positive (corresponding to that case of the generalised PSO model, in which at each time step the number of newly created internal links is larger than the number of deleted internal links), the average internal degree becomes larger as the degree threshold begins to increase.  For $L=0$ (corresponding to the case of the original PSO model) the average internal degree remains constant until the degree threshold does not become so large that the subgraphs become extremely small. And finally, for negative $L$ (corresponding to that case of the generalised PSO model, in which at each time step more internal links are deleted than created) with the increase of the degree threshold the average internal degree decreases even for relatively small values of the threshold. Note that the shape of the $\left< k_{\mathrm{internal}}\right>-k_{\rm min}$ curve does not depend on the popularity fading parameter $\beta$, thus neither on the exponent $\gamma$ of the degree distribution, as opposed to the $\mathbb{S}^1/\mathbb{H}^2$ model\cite{Boguna_2008_PRL}, where the average internal degree is an increasing function of the degree threshold only for $\gamma<3$. 

\section{Setting the embedding parameters for ncMCE, its angular optimisation and HyperMap}

When determining the radial coordinates in the ncMCE method, the popularity fading parameter $\beta$ has to be set to some specific value, whereas for the calculation of the logarithmic loss in our angular optimisation process, besides $\beta$ also the temperature $T$ and the parameter $m$ have to be specified. In the case of HyperMap, the parameter $L$ has to be inputted too.
%For determining the radial coordinates in the ncMCE method the popularity fading parameter $\beta$, while for calculating the logarithmic loss with respect to the E-PSO model in HyperMap or during our angular optimisation process also the temperature $T$ and the parameters $m$ and $L$ related to the average degree as $\left< k\right>=2(m+L)$ are needed to be specified.

Changing the value of the curvature $K=-\zeta^2$ of the hyperbolic plane corresponds to a simple rescaling of all hyperbolic distances, which does not affect how the node distances are related to each other; thus, the value of $\zeta$ practically does not affect the embedding, and -- according to the standard practice in the literature -- it can always be simply set to $1$. However, the embedding quality strongly depends on the parameters $m$, $L$, $\beta$ and $T$. The usual approach for setting these parameters relies on estimations based on observed statistical features of the graph to be embedded \cite{HyperMap}: $m$ can be set to the minimum observed node degree in the network, $L$ can be obtained from the average degree as $L=\left< k\right>/2-m$, and $\beta$ can be calculated as $\beta=1/(\gamma-1)$ from the degree decay exponent $\gamma$ determined by fitting a power-law to the degree distribution. To infer $T$ experimentally we can use the empirical connection probability\cite{HyperMap} or the average clustering coefficient of the network \cite{Alanis-Lobat_liekly_LE_emb} through a relatively complicated procedure. Nevertheless, for example fitting a power-law to the degree distribution of a network can be problematic in many cases, partly because of the uncertainty in the identification of that range of the degree distribution over which the power-law behaviour holds, and partly because for the fitting that very part of the distribution (namely the tail) has to be considered which corresponds to the rather rare events and thereby is complicated by the occurrence of large fluctuations. Therefore, instead of following the usual, frequently laborious procedures for estimating the embedding parameters, we applied a less burdensome method which is able to determine all the necessary parameters simultaneously. 

An important note is that when a network is assumed to be generated by the E-PSO model, the above-mentioned parameters characterise the adjacency matrix itself, and not a certain hyperbolic arrangement of the network. Therefore, we can assume that by optimising the parameters $m$, $L$, $\beta$ and $T$ for just one particular embedding of the examined network, we can actually get close to that parametrisation of the E-PSO model which is the most correspondent with the network in general.
%Since the parameters of the E-PSO model assumed to form the network in question characterises the adjacency matrix itself and not a certain hyperbolic arrangement of the network, therefore it can be assumed that by seeking for the optimal parameters $m$, $L$, $\beta$ and $T$ for just one given embedding of the network we actually determine that parametrisation of the E-PSO model which is the most correspondent with the network generally. 
As a consequence, the embedding parameters can be estimated for example based on a node arrangement obtained with the ncMCE method, which is a reasonable choice because of its low running time. Given a hyperbolic embedding of a network, we can consider the parameters to be optimal when the logarithmic loss $LL$ of this embedding is minimal.

According to the above, in our algorithm proposed for parameter setting we simply create an embedding of the network with ncMCE and determine the minimum point of its logarithmic loss via a gradient descent in the $m-\beta-T$ parameter space, meaning that we take ever smaller steps in the direction of the negative gradient of the logarithmic loss (i.e. the vector $(-\frac{\partial LL}{\partial m},-\frac{\partial LL}{\partial \beta},-\frac{\partial LL}{\partial T})$) until we get so close to the optimum that the resultant step size becomes smaller than a given value, or in other words, until the optimum is approached with a given precision. The parameter $L$ needed for HyperMap is calculated then from the relation $\left< k\right>=2(m+L)$ using the optimal value of $m$.

In the case of embedding synthetic networks obtained from the PSO model, we started the search in the parameter space from the parameters used for the network generation, while for real networks we used $\beta=0.5$, $T=0.5$ and $m=\left< k\right>/2$ as the starting point. In accordance with our generalised E-PSO model, $L$ was allowed to take negative values as well, i.e. $m$ was allowed to take values larger than $\left< k\right>/2$. We permitted $m$ to change between $1$ and $2\cdot\left< k\right>$ ($m$ never increased over this value even if it was not prohibited), while the value of $\beta$ and $T$ was restricted to the interval $[0.1,0.99]$. %HOGY NE SZÁLLJON EL A PROGRAM AMIATT HOGY NEM TUDJA KISZÁMOLNI AZ LL DERIVÁLTJÁT. 
If the end point of a step would have fallen outside from the designated range of any of the parameters, we set the involved parameters to their allowed extremum in this step. The step size was tuned separately in each parameter's direction. The size of the first step was set to the distance of the starting point from the permitted extremum falling in the direction of the initial negative derivative of the logarithmic loss multiplied by a constant factor smaller than $1$, where the multiplying constant was the same for all three partial derivatives and it was set experimentally to a value at which the algorithm seemed to eventually converge. %[[Megj.: Tehát a kezdőponton és a kívánt precízión kívül még azt lehet állítani, hogy az első lépésben az optimum irányában rendelkezésre álló hely hanyadrészén ugorjunk át.]] 
The size of the following steps was calculated in each direction as the corresponding partial derivative multiplied by another constant factor; thus, together with the partial derivatives, the step sizes declined as the optimum was approached. The multiplying constant used from the second step was calculated by dividing the size of the first step in the given direction by the absolute value of the corresponding initial partial derivative. This way it was provided in each direction that -- unless the size of the first step was set to a too large value which led to the increase of the partial derivative -- the size of the second step was always smaller compared to the first step.

\section{Embedding of synthetic networks generated by the PSO model}

This section presents our results concerning synthetic networks. We compared the performance of the original ncMCE\cite{coalescentEmbedding}, ncMCE with our angular optimisation, HyperMap\cite{HyperMap} and Mercator\cite{Boguna_embedding_2019} for networks generated by the PSO model with different combinations of the input parameters. The effect of changing the number of nodes $N$ is shown in the main article, while the dependence of the embedding quality measures on the expected average degree $2m$, the popularity fading parameter $\beta$ and the temperature $T$ used for the network generation is shown here below. An important remark is that both the logarithmic loss\cite{HyperMap} $LL$ and the greedy routing score\cite{coalescentEmbedding} $GR$ seem to converge after only a few rounds of angular optimisation for all types of synthetic networks that we studied, which is indispensable for keeping the running time of the angular optimisation low.

Besides, we demonstrate through the embedding of synthetic networks that for all the four investigated embedding methods the repeated embedding of the same network leads to hyperbolic arrangements of different quality. According to our experiments, the distribution of the embedding quality obtained by the repetition of a given method is a bell-shaped curve or consists of more bell-shaped peaks. Knowing this, we show how the improvement in both the logarithmic loss $LL$ and the greedy routing score $GR$ can be estimated after some trials as a function of the number of repetitions. The impact of the network generation parameters $N$, $m$, $\beta$ and $T$ on the variance of the embedding quality occurring during the repetition of the embedding is also investigated below.

\subsection{The effect of the network generation parameters on the embedding quality}

\begin{figure}
    \centering
    \includegraphics[width=\textwidth]{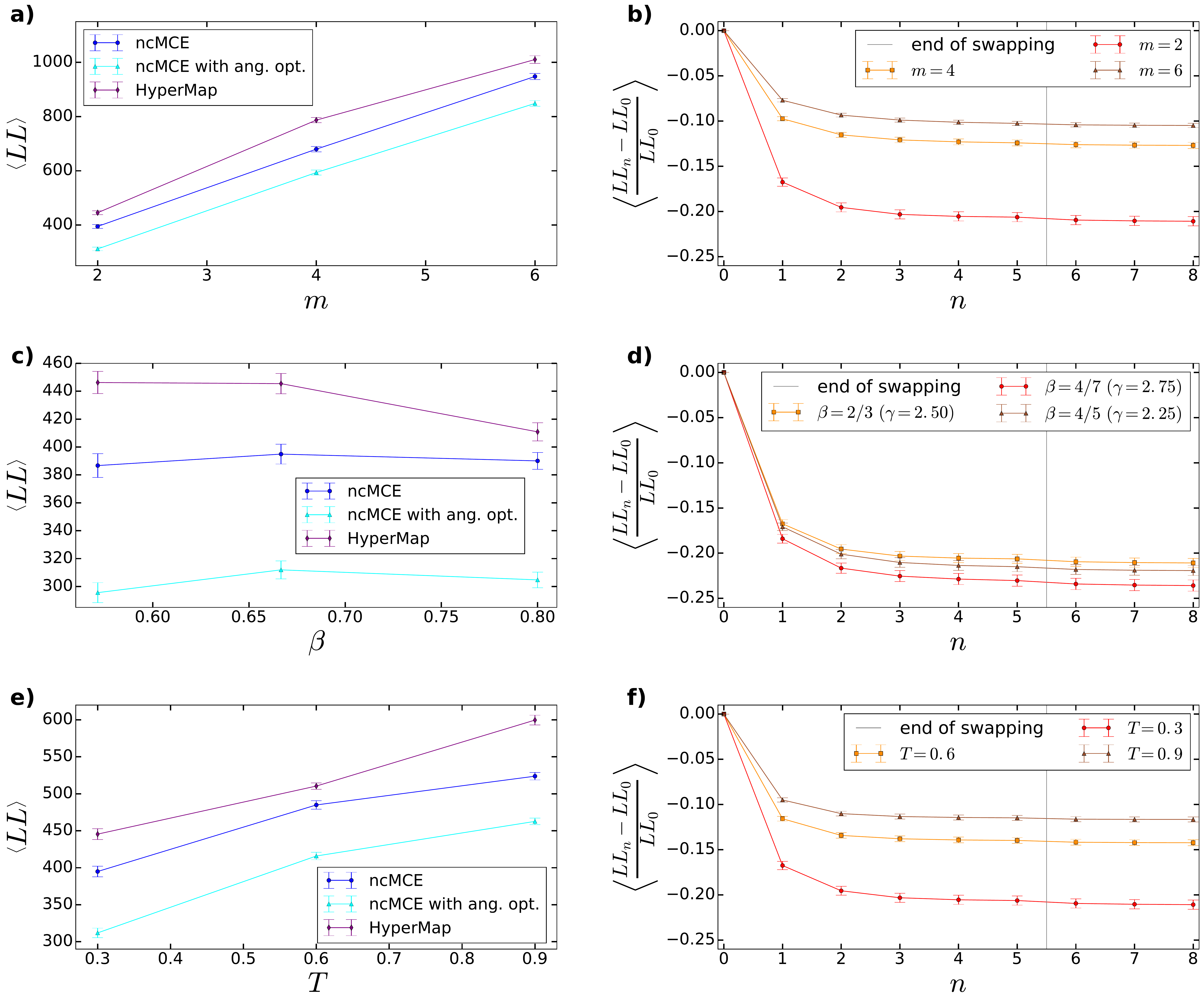}
    \caption{{\bf The impact of the network generation parameters on the logarithmic loss achieved by the original ncMCE, ncMCE with angular optimisation and HyperMap for networks generated by the PSO model.} Logarithmic loss $LL$ obtained with the studied embedding methods as a function of network generation parameters (left) and the relative improvement in the logarithmic loss as a function of the number of rounds $n$ during the angular optimisation of the node arrangement resulted from the ncMCE method (right). Each data point corresponds to a value averaged over $100$ synthetic networks, the bars indicate the $95\%$ confidence intervals. a) $LL$ as a function of the half of expected average degree $m$ when $\zeta=1$, $N=100$, $\beta=2/3$ and $T=0.3$. b) Convergence of $LL$ as a function of $n$ under the same settings as in panel a). c) $LL$ as a function of the popularity fading parameter $\beta$ when $\zeta=1$, $N=100$, $m=2$ and $T=0.3$. d) The convergence of $LL$ as a function of $n$ under the same settings as in panel c). e) $LL$ as a function of the temperature $T$ when $\zeta=1$, $N=100$, $m=2$ and $\beta=2/3$. f) The convergence of $LL$ as a function of $n$ under the same settings as in panel e).}
    \label{fig:synthLL}
\end{figure}

\begin{figure}
    \centering
    \includegraphics[width=\textwidth]{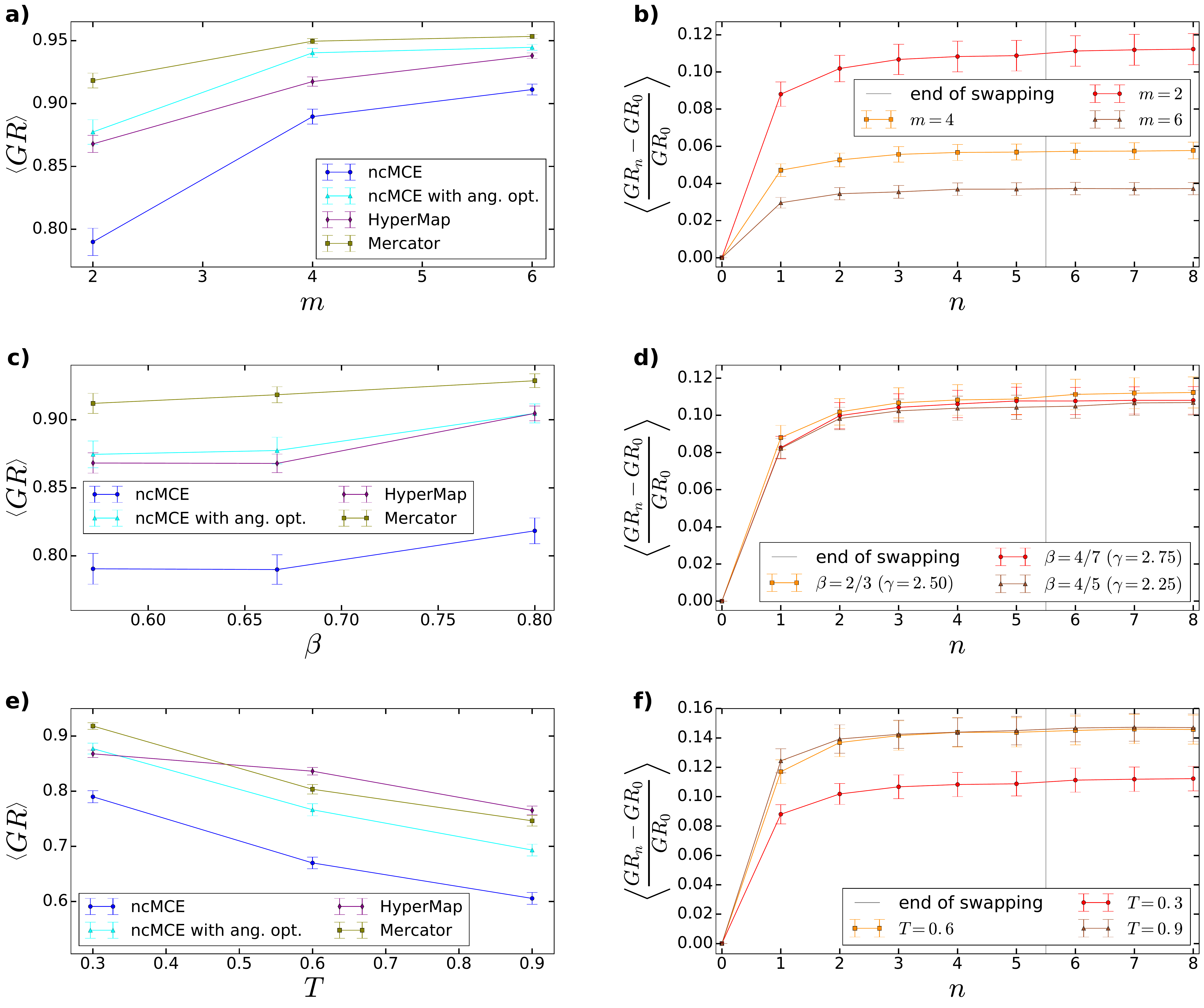}
    \caption{{\bf The impact of the network generation parameters on the greedy routing score achieved by the original ncMCE, ncMCE with angular optimisation, HyperMap and Mercator for networks generated by the PSO model.} Greedy routing score $GR$ obtained with the studied embedding methods as a function of network generation parameters (left) and the relative improvement in the greedy routing score as a function of the number of rounds $n$ during the angular optimisation of the node arrangement resulted from the ncMCE method (right). Each data point corresponds to a value averaged over $100$ synthetic networks, the bars indicate the $95\%$ confidence intervals. a) $GR$ as a function of the half of expected average degree $m$ when $\zeta=1$, $N=100$, $\beta=2/3$ and $T=0.3$. b) Convergence of $GR$ as a function of $n$ under the same settings as in panel a). c) $GR$ as a function of the popularity fading parameter $\beta$ when $\zeta=1$, $N=100$, $m=2$ and $T=0.3$. d) The convergence of $GR$ as a function of $n$ under the same settings as in panel c). e) $GR$ as a function of the temperature $T$ when $\zeta=1$, $N=100$, $m=2$ and $\beta=2/3$. f) The convergence of $GR$ as a function of $n$ under the same settings as in panel e).}
    \label{fig:synthGR}
\end{figure}

We tested four embedding methods (namely ncMCE, ncMCE with our angular optimisation, HyperMap and Mercator) on networks generated by the PSO model using different parameter combinations. The dependence of the logarithmic loss $LL$ and the greedy routing score $GR$ on the number of nodes $N$ is depicted in the main article, and here in Figs. \ref{fig:synthLL}a, \ref{fig:synthLL}c, \ref{fig:synthLL}e and Figs. \ref{fig:synthGR}a, \ref{fig:synthGR}c, \ref{fig:synthGR}e we present the dependence of these quality measures on the half of expected average degree $m$, the popularity fading parameter $\beta$ (or the expected exponent of the degree distribution $\gamma=1+1/\beta$) and the temperature $T$ (of which the expected average clustering coefficient is a decreasing function) used for network generation. A number of $100$ networks were generated for each parameter setting. We then regarded the model parameters as unknown, and set them for ncMCE, its optimisation and HyperMap according to the parameter optimisation method described in the previous section. In the case of Mercator, we used its own parameter estimation process included in the algorithm \cite{Boguna_embedding_2019}. Each network was embedded once with each embedding method and the obtained quality measures were averaged over the networks of the same generation parameters. For each network we set two radial orders of the nodes: one for the parameter optimisation and -- in order not to favour ncMCE over the other methods -- another with which the performance of the ncMCE method was measured. This latter radial order was identical to the radial order used in our angular optimisation process and the radial order used in HyperMap. (Note that Mercator does not use a strict radial ordering among the network nodes.) According to Figs. \ref{fig:synthLL}b, \ref{fig:synthLL}d, \ref{fig:synthLL}f and Figs. \ref{fig:synthGR}b, \ref{fig:synthGR}d, \ref{fig:synthGR}f, after only $5$ swapping and $3$ non-swapping rounds of angular optimisation both the logarithmic loss $LL$ and the greedy routing score $GR$ reach a steady value for all type of PSO networks, which is in the case of $LL$ $10-20\%$, while for $GR$ $4-15\%$ better than the result of ncMCE without optimisation.

\subsection{Varying quality scores under the repetition of the embedding}

This section sheds light on the fact that for all four embedding methods that we examined, the hyperbolic arrangements resulting from the repeated execution of the embedding may not be equivalent regarding their quality scores. We generated networks with the PSO model and embedded them repeatedly with the original ncMCE, ncMCE with angular optimisation, HyperMap and Mercator. In the former three cases the quality differences between the repetitions are indicated by the changes both in the logarithmic loss $LL$ and the greedy routing score $GR$, and the performance of Mercator -- measured only by the greedy routing score $GR$ -- also varied under the repetitions.

The ncMCE method determines the radial coordinates of the network nodes in the logarithmic loss optimising way proposed originally as a part of the HyperMap method. As detailed in the main article, in order to minimise the logarithmic loss of the embedding with respect to the E-PSO model, HyperMap assigns logarithmically increasing radial coordinates to the nodes in the decreasing order of the node degrees. However, the radial order among the nodes of the same degree can not be determined analytically, and according to the commonly applied procedure the ties in the degree order can be broken arbitrarily. Due to this ambiguity in the radial ordering of the nodes, for the same network usually many different arrangements of the network nodes can be produced by HyperMap, the original ncMCE and ncMCE with angular optimisation. To study the effect of the allowed changes in the radial order of the nodes we embedded PSO networks with the aforementioned three methods multiple times, where the embeddings of a given network were carried out using in each repetition the same setting of the embedding parameters $\zeta$, $m$, $L$, $\beta$ and $T$ (obtained with the logarithmic loss minimising method described in a previous section), but a randomly chosen permutation of the radial order of nodes with equal degree. Mercator does not allow the specification of the radial order; thus, this algorithm was just simply re-run from scratch.

\begin{figure}
    \centering
    \includegraphics[width=\textwidth]{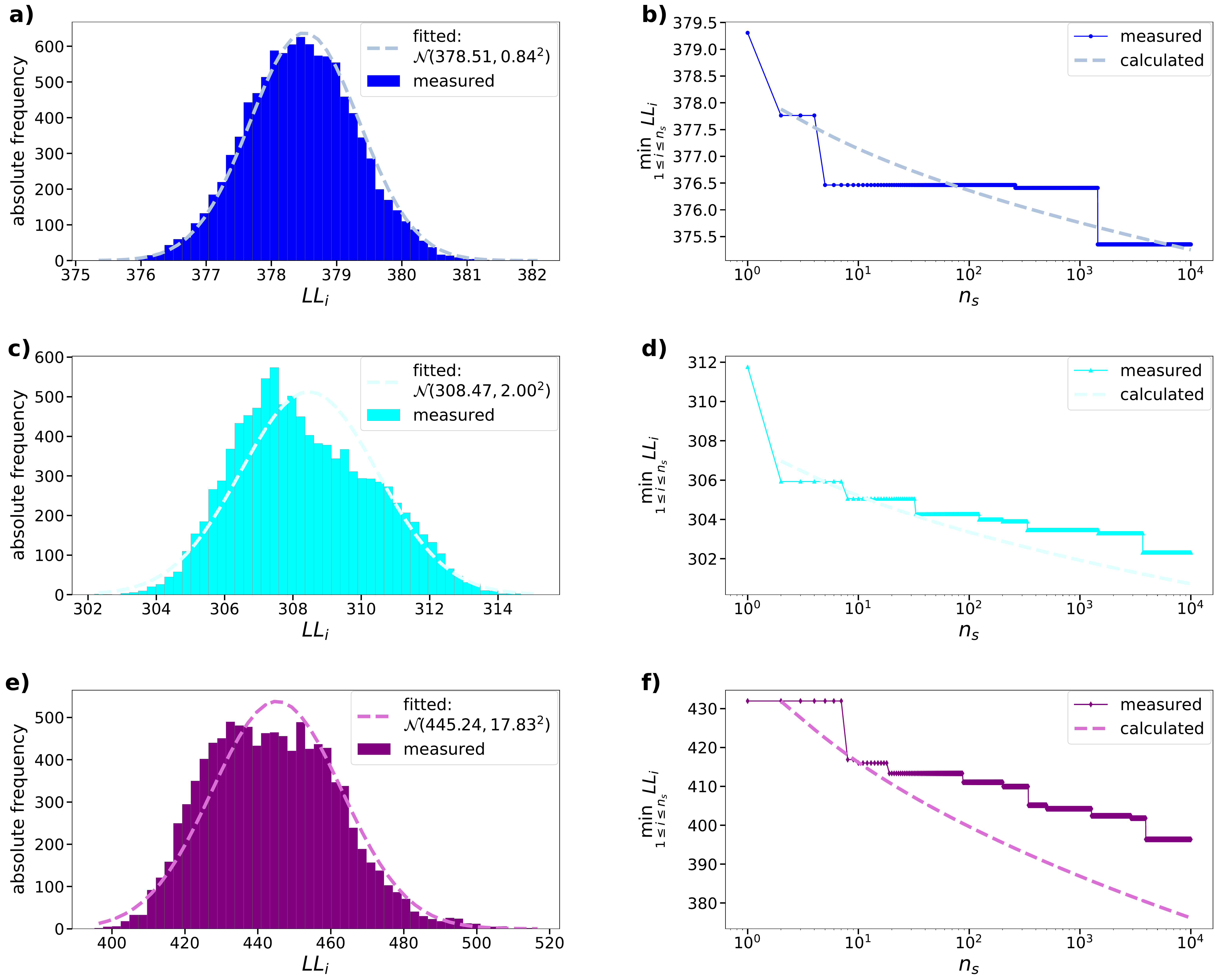}
    \caption{{\bf The distribution of the logarithmic loss $LL$ among the repeated embeddings of a network generated by the PSO model parametrised by $\zeta=1$, $N=100$, $m=2$, $\beta=2/3$ and $T=0.3$.} In each row of the figure the results regarding one of the studied embedding algorithms are presented: panels a) and b) refer to ncMCE, panels c) and d) refer to ncMCE with our angular optimisation and panels e) and f) refer to HyperMap. The left panels show the observed quality distributions and the normal distributions fitted to these data, while the right panels depict the achieved best logarithmic loss as a function of the number of repetitions of the embedding. The dashed curves on the right are obtained from equation (\ref{eq:LL_extr}) by substituting in the mean and the standard deviation of the corresponding fitted normal distribution on the left.}
    \label{fig:synthHistLL}
\end{figure}

\begin{figure}
    \centering
    \includegraphics[width=\textwidth]{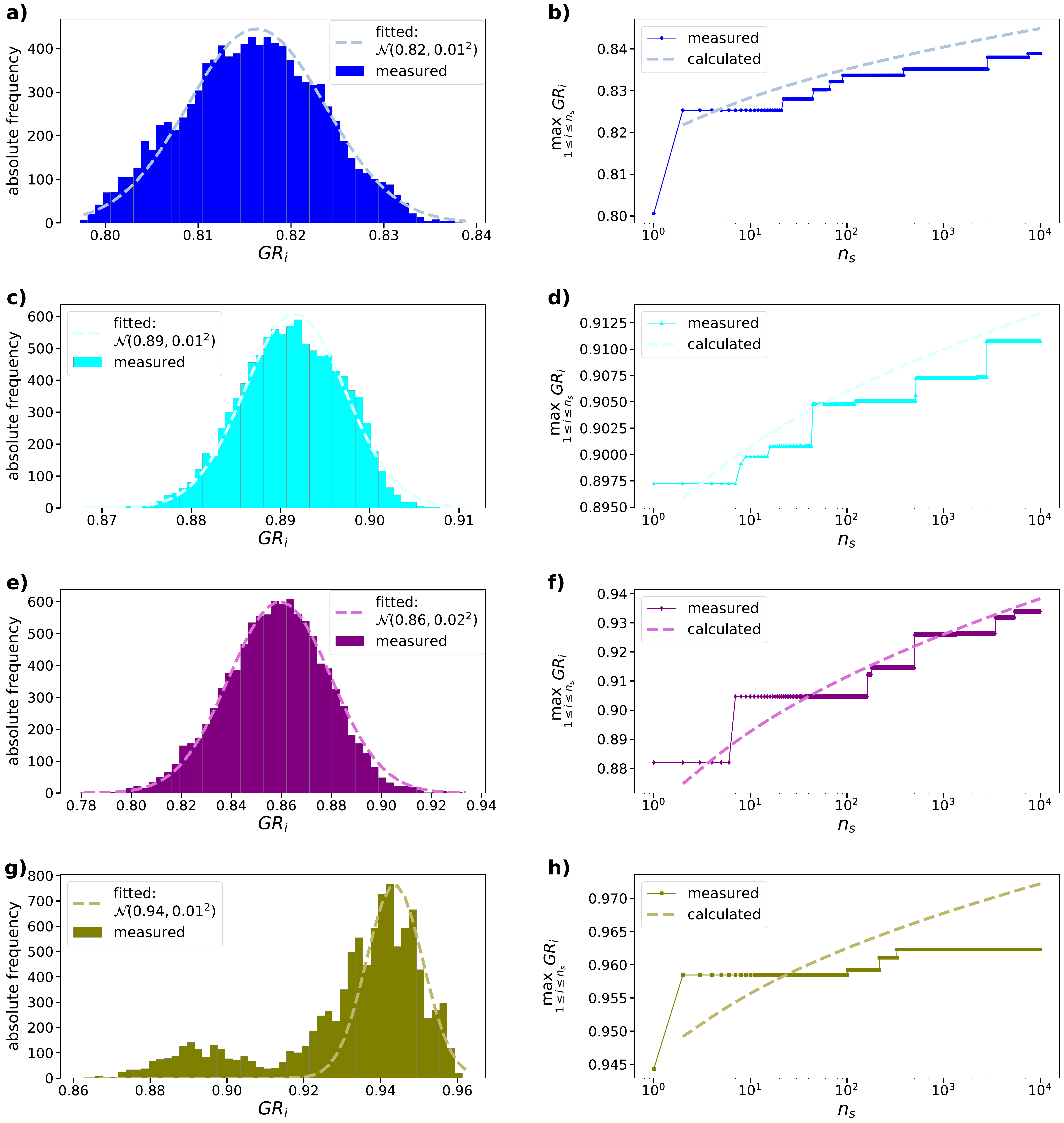}
    \caption{{\bf The distribution of the greedy routing score $GR$ among the repeated embeddings of a network generated by the PSO model parametrised by $\zeta=1$, $N=100$, $m=2$, $\beta=2/3$ and $T=0.3$.} In each row of the figure the results regarding one of the studied embedding algorithms are presented: panels a) and b) refer to ncMCE, panels c) and d) refer to ncMCE with our angular optimisation, panels e) and f) refer to HyperMap and panels g) and h) refer to Mercator. The left panels show the observed quality distributions and the normal distributions fitted to these data, while the right panels depict the achieved best greedy routing score as a function of the number of repetitions of the embedding. The dashed curves on the right are obtained from equation (\ref{eq:GR_extr}) by substituting in the mean and the standard deviation of the corresponding fitted normal distribution on the left. %[Mivel a Mercatornál 2 csúcs lett és csak a jobb értékeknél lévőre akartam illeszteni, találomra azt mondtam, hogy csak 0.93 felett illesztek (és az ide eső össz előfordulási szám*binszélesség-gel szorzom az 1-re normált pdf-et)].
    }
    \label{fig:synthHistGR}
\end{figure}

Figures \ref{fig:synthHistLL}a, \ref{fig:synthHistLL}c, \ref{fig:synthHistLL}e and Figures \ref{fig:synthHistGR}a, \ref{fig:synthHistLL}c, \ref{fig:synthHistLL}e exemplify that for synthetic networks generated by the PSO model the distribution of the embedding quality among the repetitions of the embedding is close to a Gaussian for the original ncMCE, ncMCE with angular optimisation and HyperMap. Note that the original ncMCE method is fundamentally different from its angularly optimised version and HyperMap in that respect, that in the former the angular coordinates are determined independently from the radial arrangement of the nodes, while in the latter two methods the angular arrangement depends on the actual radial node order. Thereby, it can be expected that the distribution of the embedding quality among the different radial orderings of the nodes is Gaussian for any embedding method which determines the radial coordinates the same way as HyperMap. Figure \ref{fig:synthHistGR}g demonstrates that the $GR$ distribution obtained from the repeated execution of Mercator is not a simple normal distribution, but instead it consists of more bell-shaped peaks. However, if we are interested only in the best results achievable by repeating the embedding, we have to take into consideration only the peak at the end of the largest $GR$ values, which in itself is similar to a normal distribution.

Assuming that the distribution of the quality score follows a normal distribution ${\mathcal N}(\mu,\sigma)$ (where $\mu$ is the mean and $\sigma$ corresponds to the standard deviation), we can obtain an estimate for the best score that can be achieved under $n_s$ number of repetitions of the embedding.
%Due to the bell-shaped peaks in the quality distributions, based on some trials the embedding quality attainable by accomplishing a given number $n_s$ of repetitions of the embedding can be estimated.
The expected value for the extreme values of our interest among $n_s$ number of samples can be formulated as
\begin{subequations}
\begin{align}
     \mathbb{E}\bigg[\min_{1\leq i\leq n_s}\it{LL}_{\it{i}}\bigg] &=\rm\mu_{\it{LL}} - \sigma_{\it{LL}}\cdot g(n_s), \label{eq:LL_extr}\\ \mathbb{E}\bigg[\max_{1\leq i\leq n_s}\it{GR}_{\it{i}}\bigg] &=\rm\mu_{\it{GR}} + \sigma_{\it{GR}}\cdot g(n_s), \label{eq:GR_extr}
\end{align}
\end{subequations}
where $n_s$ is assumed to be $n_s\geq 2$ and the function $g(n_s)$ is given by
\begin{equation}
    g(n_s)=\sqrt{2\mathrm{ln}(n_s)}-\frac{\mathrm{ln}(\mathrm{ln}(n_s))+\mathrm{ln}(4\pi)-2\Gamma}{2\sqrt{2\mathrm{ln}(n_s)}}+\mathcal{O}\bigg(\frac{1}{\mathrm{ln}(n_s)}\bigg),
    \label{eq:fittedFunctionsForMoreTrials}
\end{equation}
where $\Gamma=0.5772156649...$ is the Euler–Mascheroni constant\cite{GaussExpectedMin}.
%For a normal distribution with mean $\mu$ and variance $\sigma^2$ the expected value of the first value from the bottom/top in a sample of $n_s$ elements can be calculated according to equation (\ref{eq:fittedFunctionsForMoreTrials}), where $\Gamma=0.5772156649...$ is the Euler–Mascheroni constant\cite{GaussExpectedMin}. 
We fitted a normal distribution to each of the measured quality score distributions (in the case of Mercator only to the relevant part of the distribution) and substituted the obtained $\mu$ and $\sigma$ parameters into equations (\ref{eq:LL_extr}-\ref{eq:fittedFunctionsForMoreTrials}). According to Figs. \ref{fig:synthHistLL}b, \ref{fig:synthHistLL}d, \ref{fig:synthHistLL}f and Figs. \ref{fig:synthHistGR}.b, \ref{fig:synthHistGR}d, \ref{fig:synthHistGR}f, \ref{fig:synthHistGR}h, the resulting functions are close to the curves describing the measured best quality score as a function of the number of embeddings carried out. This suggests that the improvement in the logarithmic loss or in the greedy routing score achievable by further repetitions of any of the four studied embedding methods can be predicted by simply fitting equations (\ref{eq:LL_extr}-\ref{eq:fittedFunctionsForMoreTrials}) to the lowest achieved logarithmic loss or the highest achieved greedy routing score as a function of the number of trials so far.

%\begin{equation}
%  \begin{gathered}
%    \mathbb{E}\bigg[\min_{1\leq i\leq n_s}\it{LL}_{\it{i}}\bigg] =\rm\mu_{\it{LL}} - \sigma_{\it{LL}}\cdot g(n_s) \text{ and } \mathbb{E}\bigg[\max_{1\leq i\leq n_s}\it{GR}_{\it{i}}\bigg] =\rm\mu_{\it{GR}} + \sigma_{\it{GR}}\cdot g(n_s) \text{ for } 2\leq n_s \text{,}\\
%    \text{where } g(n_s)=\sqrt{2\mathrm{ln}(n_s)}-\frac{\mathrm{ln}(\mathrm{ln}(n_s))+\mathrm{ln}(4\pi)-2\Gamma}{2\sqrt{2\mathrm{ln}(n_s)}}+\mathcal{O}\bigg(\frac{1}{\mathrm{ln}(n_s)}\bigg)
%  \end{gathered}
%  \label{eq:fittedFunctionsForMoreTrials}
%\end{equation}

\begin{figure}
    \centering
    \includegraphics[width=\textwidth]{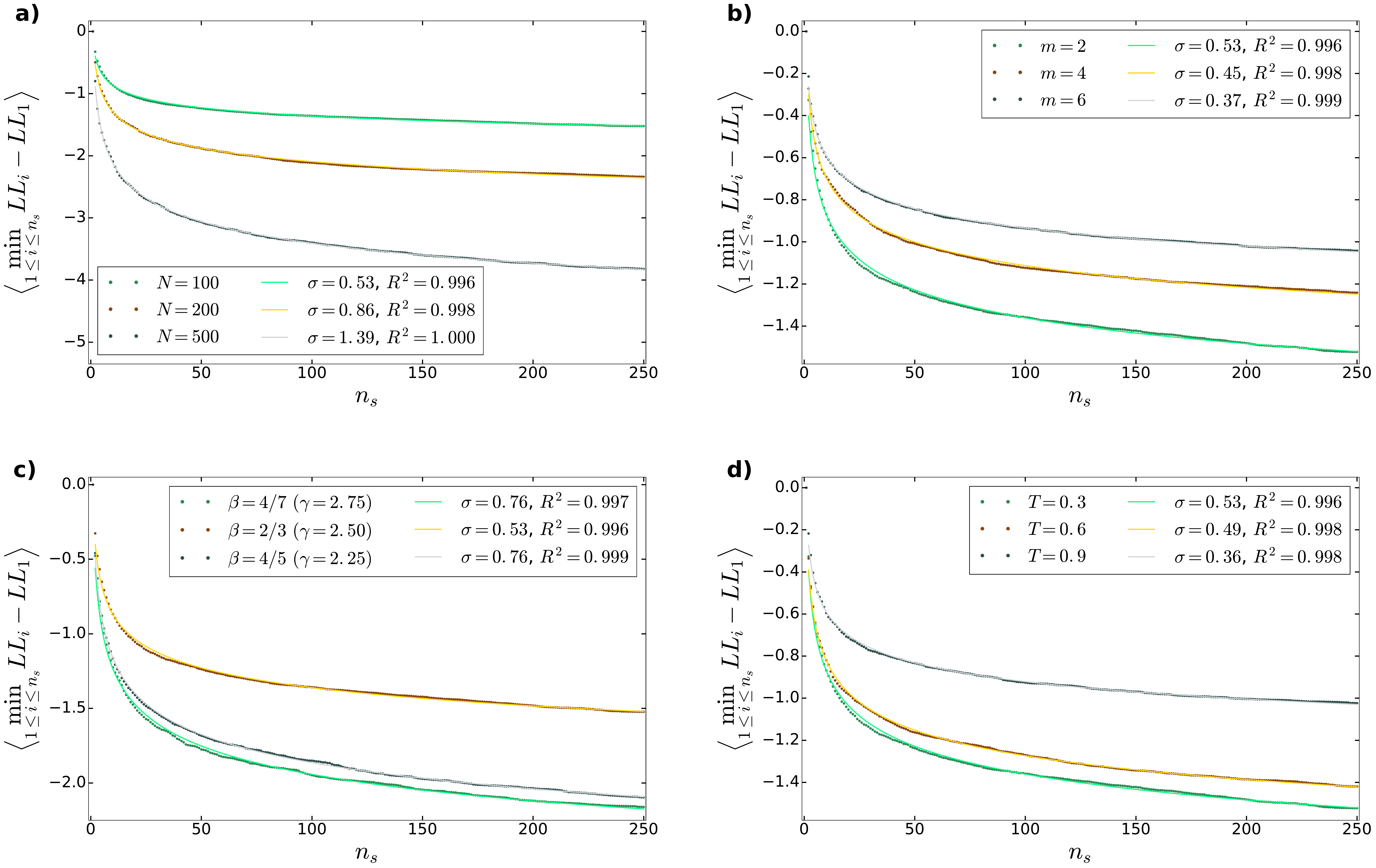}
    \caption{{\bf The impact of the network generation parameters on the variance of the logarithmic loss $LL$ observed during the repeated execution of ncMCE for networks generated by the PSO model.} The depicted data points correspond to the improvement in the embedding quality averaged over $500$ PSO networks generated using a given parameter setting. The solid lines are obtained by fitting equation (\ref{eq:LL_extr}) to the measured curves. The fitted coefficient is the standard deviation $\sigma$ characterising the distribution of the improvement in the logarithmic loss compared to the first trial of the embedding. Note that since not the absolute values, but the improvements are plotted, the mean of the quality distribution is always $0$. The coefficient of determination $R^2$ is also given for each fit in the legends. a) The improvement in $LL$ as a function of the number of repetitions of the embedding at different $N$ parameters with fixed $\zeta=1$, $m=2$, $\beta=2/3$ and $T=0.3$. b) The improvement in $LL$ as a function of the number of repetitions of the embedding at different $m$ parameters with fixed $\zeta=1$, $N=100$, $\beta=2/3$ and $T=0.3$. c) The improvement in $LL$ as a function of the number of repetitions of the embedding at different $\beta$ parameters with fixed $\zeta=1$, $N=100$, $m=2$ and $T=0.3$. d) The improvement in $LL$ as a function of the number of repetitions of the embedding at different $T$ parameters with fixed $\zeta=1$, $N=100$, $m=2$ and $\beta=2/3$.}
    \label{fig:radOrder_ncMCE_LL}
\end{figure}

\begin{figure}
    \centering
    \includegraphics[width=\textwidth]{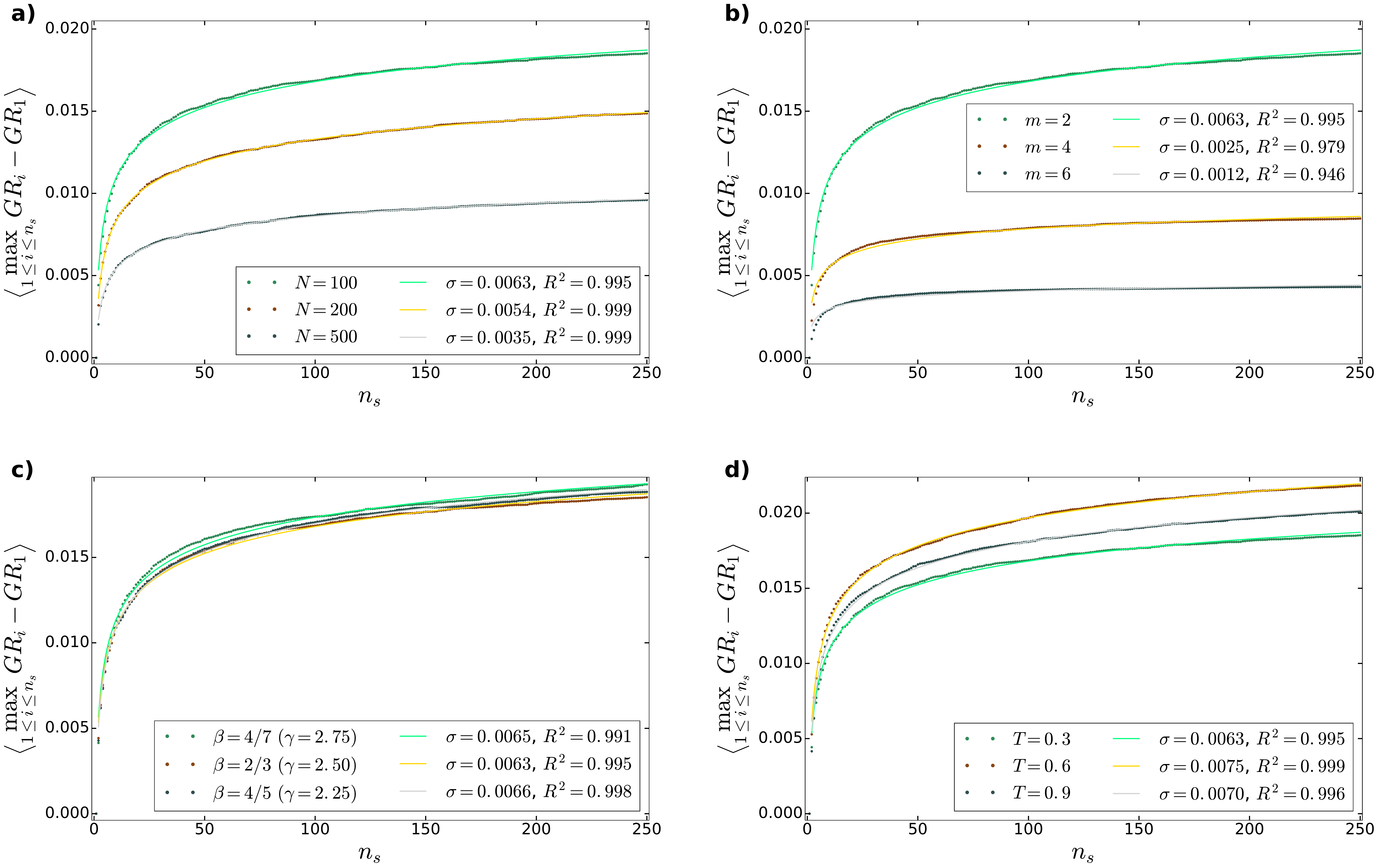}
    \caption{{\bf The impact of the network generation parameters on the variance of the greedy routing score $GR$ observed during the repeated execution of ncMCE for networks generated by the PSO model.} The depicted data points correspond to the improvement in the embedding quality averaged over $500$ PSO networks generated using a given parameter setting. The solid lines are obtained by fitting equation (\ref{eq:GR_extr}) to the measured curves. The fitted coefficient is the standard deviation $\sigma$ characterising the distribution of the improvement in the greedy routing score compared to the first trial of the embedding. Note that since not the absolute values, but the improvements are plotted, the mean of the quality distribution is always $0$. The coefficient of determination $R^2$ is also given for each fit in the legends. a) The improvement in $GR$ as a function of the number of repetitions of the embedding at different $N$ parameters with fixed $\zeta=1$, $m=2$, $\beta=2/3$ and $T=0.3$. b) The improvement in $GR$ as a function of the number of repetitions of the embedding at different $m$ parameters and fixed  $\zeta=1$, $N=100$, $\beta=2/3$ and $T=0.3$. c) The improvement in $GR$ as a function of the number of repetitions of the embedding at different $\beta$ parameters and fixed  $\zeta=1$, $N=100$, $m=2$ and $T=0.3$. d) The improvement in $GR$ as a function of the number of repetitions of the embedding at different $T$ parameters and fixed $\zeta=1$, $N=100$, $m=2$ and $\beta=2/3$.}
    \label{fig:radOrder_ncMCE_GR}
\end{figure}

\begin{figure}
    \centering
    \includegraphics[width=\textwidth]{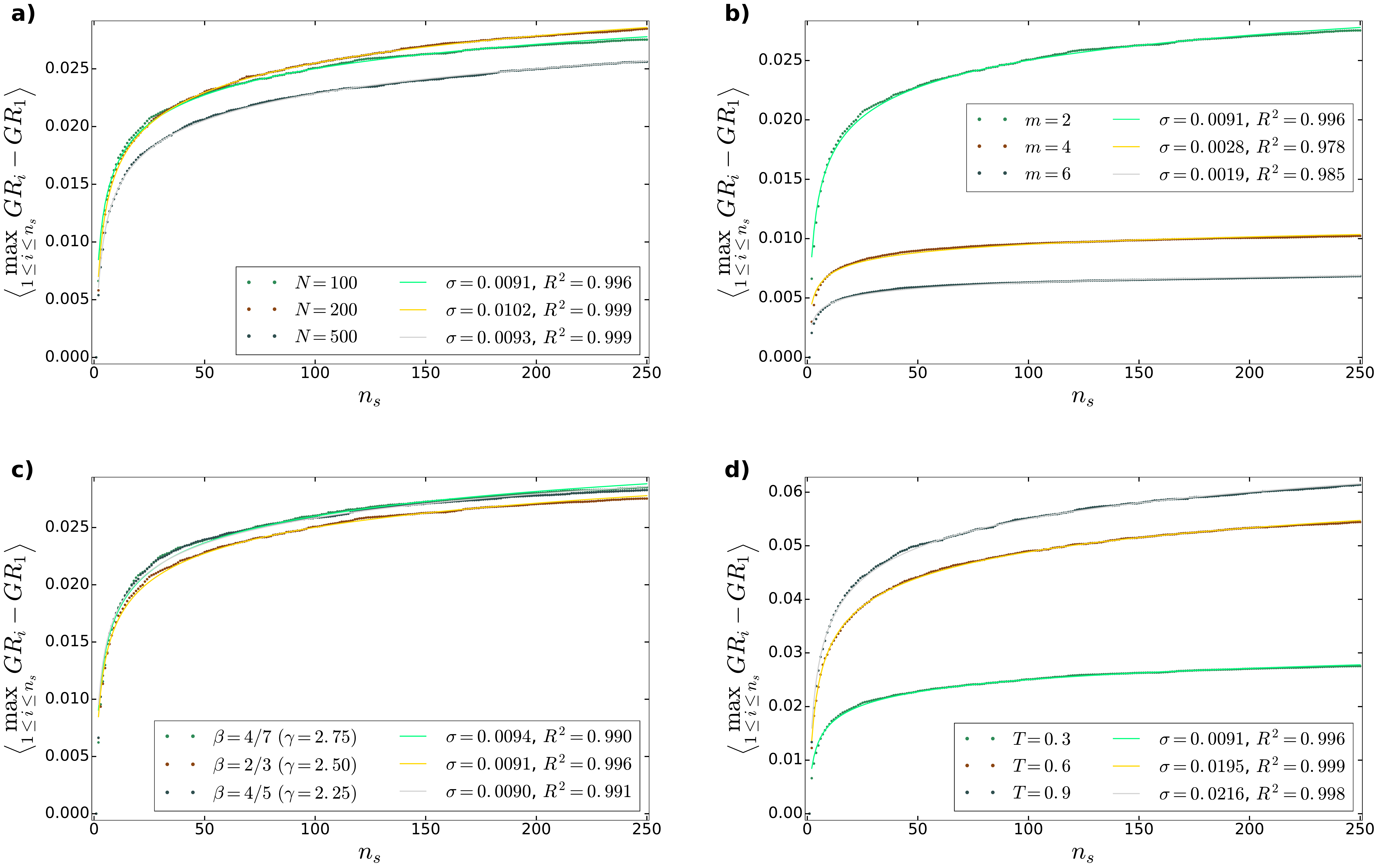}
    \caption{{\bf The impact of the network generation parameters on the variance of the greedy routing score $GR$ observed during the repeated execution of Mercator for networks generated by the PSO model.} The depicted data points correspond to the improvement in the embedding quality averaged over $500$ PSO networks generated using a given parameter setting. The solid lines are obtained by fitting equation (\ref{eq:GR_extr}) to the measured curves. The fitted coefficient is the standard deviation $\sigma$ characterising the distribution of the improvement in the greedy routing score compared to the first trial of the embedding. Note that since not the absolute values, but the improvements are plotted, the mean of the quality distribution is always $0$. The coefficient of determination $R^2$ is also given for each fit in the legends. a) The improvement in $GR$ as a function of the number of repetitions of the embedding at different $N$ parameters with fixed $\zeta=1$, $m=2$, $\beta=2/3$ and $T=0.3$. b) The improvement in $GR$ as a function of the number of repetitions of the embedding at different $m$ parameters and fixed  $\zeta=1$, $N=100$, $\beta=2/3$ and $T=0.3$. c) The improvement in $GR$ as a function of the number of repetitions of the embedding at different $\beta$ parameters and fixed  $\zeta=1$, $N=100$, $m=2$ and $T=0.3$. d) The improvement in $GR$ as a function of the number of repetitions of the embedding at different $T$ parameters and fixed $\zeta=1$, $N=100$, $m=2$ and $\beta=2/3$.}
    \label{fig:radOrder_Mercator_GR}
\end{figure}

We demonstrate through the original ncMCE method how the network properties can affect the variance of the embedding quality appearing when different radial orders of the nodes are tried out. We generated $500$ networks with the PSO model using different parametrisations and embedded each network $250$ times with ncMCE. The required embedding parameter $\beta$ was determined for each network once by the above described $m-\beta-T$ optimising method. During the repeated embedding of the same network only the radial order between the nodes having the same degree was allowed to change. Figures \ref{fig:radOrder_ncMCE_LL} and \ref{fig:radOrder_ncMCE_GR} show the results regarding the logarithmic loss and the greedy routing score of the embeddings, respectively. Besides, these PSO networks were also embedded $250$ times with Mercator, the only one of the four studied embedding methods which does not set a strict radial order among the network nodes based on their degree. Figure \ref{fig:radOrder_Mercator_GR} presents the improvement in the performance of Mercator due to its repeated execution in the case of several different parametrisations of the network generation.

\section{Embedding of real networks}

This section details our results related to real networks. We begin with the description of the adjustment of the embedding parameters, which is followed by the results obtained for the quality scores at different parameter settings. In addition, we show how the repetition of the embedding procedure can improve the achieved quality scores. Finally, some layouts of the networks in the native representation of the hyperbolic plane are also presented.

%After describing the different methods we tried out for setting the embedding parameters for ncMCE, its angular optimisation and HyperMap, the embedding quality obtained with each parameter setting and the effect of repeating the embeddings are presented for all of the studied networks. Lastly, examples for the spatial separation of the communities are shown on the native representation of the hyperbolic plane.

\subsection{Parameter settings for the studied networks}

\begin{figure}
    \centering
    \includegraphics[width=\textwidth]{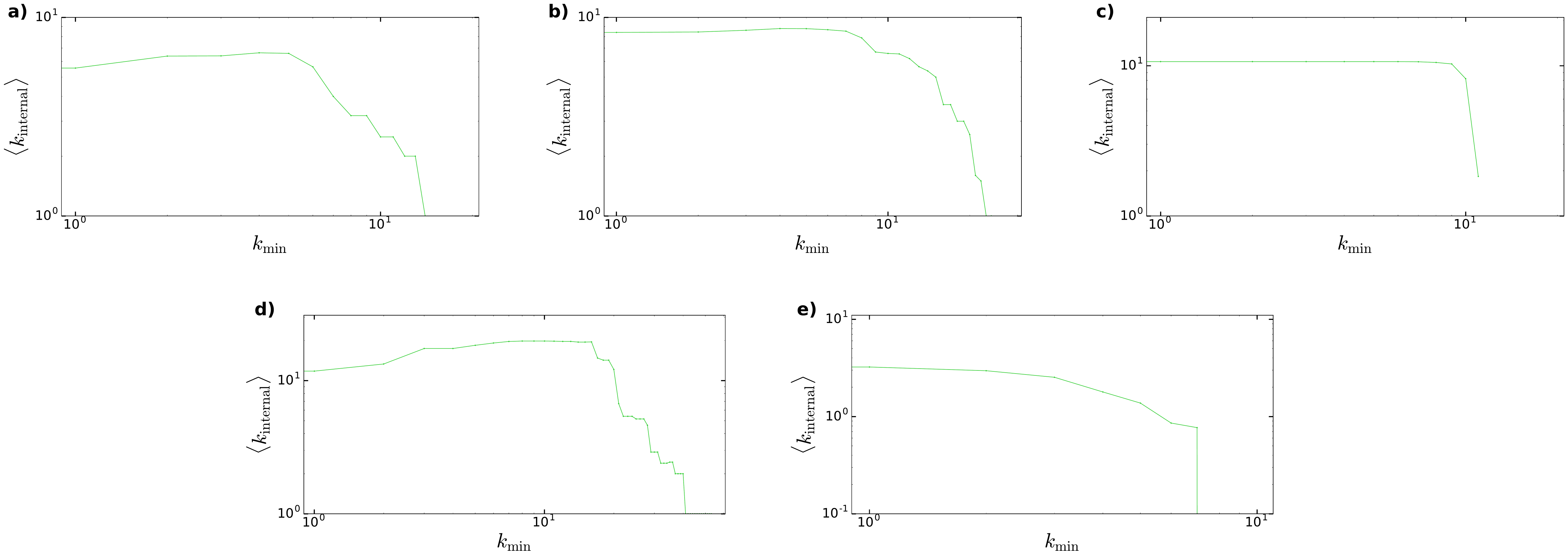}
    \caption{{\bf Average internal degree of the subgraph spanning between nodes having a degree $k>k_{\mathrm{min}}$ as a function of the degree threshold $k_{\mathrm{min}}$ for the studied real networks.} a) The twelfth layer of the Pierre Auger collaboration network. b) Network between books about US politics. c) American College Football network. d) Cambrian food web from the Burgess Shale. e) Protein interaction network from the PDZBase database.}
    \label{fig:realAvDegCurve}
\end{figure}

We studied the following real networks:
\begin{itemize}
    \item the twelfth layer of the multiplex Pierre Auger collaboration network\cite{Pierre_Auger_data}, which describes the collaborations related to the SD-reconstruction;
    \item a network\cite{pol_books_data} between books about US politics published close to the 2004 U.S. presidential election, where links correspond to frequent co-purchasing by the same buyers and nodes have been given values "l", "n", or "c" to indicate whether they are "liberal", "neutral", or "conservative";
    \item the American College Football network\cite{football_net_data} describing the games between Division IA colleges during regular season Fall 2000, where the nodes have values that indicate to which conferences they belong;
    \item a Cambrian food web from the Burgess Shale\cite{foodweb} with nodes categorised according to their trophic roles, which is in fact a directed network where links are pointing from consumers to resources, but in order to enable the hyperbolic embedding, the directedness of the connections can be disregarded;
    \item a protein interaction network\cite{pdz_base_net} from the PDZBase database.
\end{itemize}
Their characteristics related to the parameters of the logarithmic loss, namely the number of nodes $N$, the average degree $\left< k\right>$, the smallest occurring degree $\min\limits_{1\leq i\leq N} k_i$ and the average clustering coefficient $\left< c\right>$ are listed in Table \ref{table_realParameters}. 
Because of the small network sizes, fitting the degree decay exponent $\gamma$ is ambiguous in most cases; therefore, $\gamma$ values are not provided. As it was pointed out in the above section devoted to the E-PSO model, the parameter $L$ is related to the shape of the curve describing the connection between a degree threshold and the average internal degree of the subgraph determined by this threshold. Figure \ref{fig:realAvDegCurve} displays this curve for the studied real networks.

\begin{figure}
    \centering
    \includegraphics[width=\textwidth]{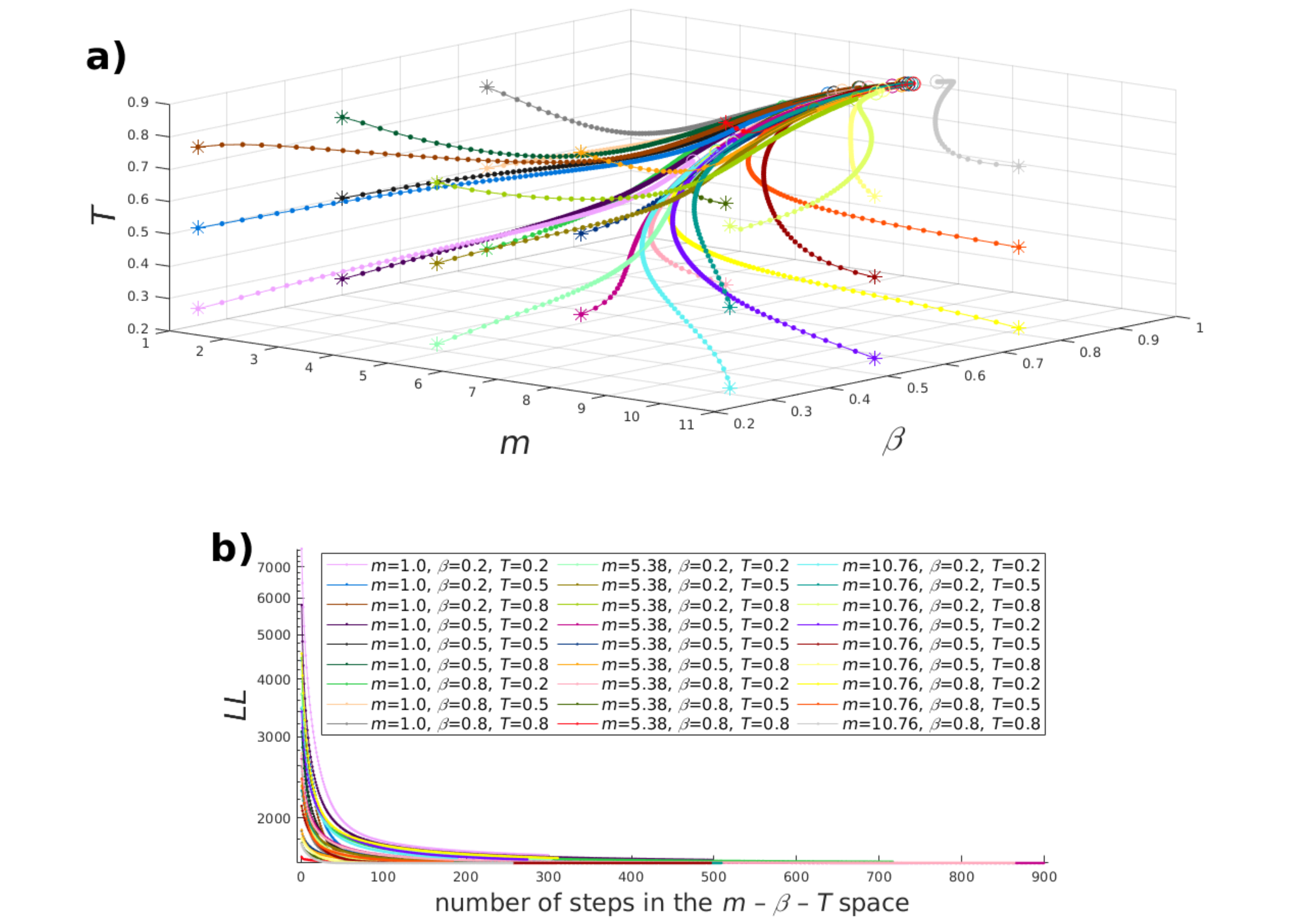}
    \caption{{\bf The operation of our method  for searching the optimal embedding parameters exemplified through the Cambrian food web from the Burgess Shale.} a) Examples of the trajectories emerging during the gradient descent used for minimising the logarithmic loss $LL$ of an ncMCE embedding in the case of different starting points in the $m-\beta-T$ parameter space. The termination points (indicated by "o") lie close to each other for all of the tested starting points (marked with "*"), suggesting that a global optimum of the logarithmic loss does exist. %[A leállás mindig akkor következett be, amikor épp megtettük az első olyan elmozdulást a paramétertérben, amelynek eredő nagysága (Euclidean norm) már nem haladta meg a 0.001-et, az 1. lépéshez tartozó, a három paraméterre közös konstans szorzó faktor pedig 0.025 volt.] 
    b) The logarithmic loss $LL$ as a function of the number of steps taken along the trajectories starting from different points of the $m-\beta-T$ space shown in panel a).}
    \label{fig:paramOptim_foodWeb}
\end{figure}

At least two different settings of the embedding parameters were tested for each real network in the case of ncMCE, its angular optimisation and HyperMap. First, the same procedure was carried out as in the case of synthetic networks: the parameters $m$, $\beta$ and $T$ were determined simultaneously by minimising the logarithmic loss $LL$ of an embedding resulted from the ncMCE method and $L$ was calculated as $\left< k\right>/2-m$. An example for the search in the 3-dimensional $m-\beta-T$ parameter space (starting from numerous different initial positions) is shown in Fig. \ref{fig:paramOptim_foodWeb}. In the second and third cases of the parameter setting, the parameters $\beta$ and $T$ were optimised in a similar way as in the first case, but here the parameter $m$ was previously fixed to $\left< k\right>/2$ and $\min\limits_{1\leq i\leq N} k_i$, respectively; thus, instead of the 3-dimensional gradient $(\frac{\partial LL}{\partial m},\frac{\partial LL}{\partial \beta},\frac{\partial LL}{\partial T})$ the 2-dimensional vector $(\frac{\partial LL}{\partial \beta},\frac{\partial LL}{\partial T})$ was used. Setting $m$ to $\left< k\right>/2$ corresponds to the assumption that the network was generated by the original PSO model, while the choice $m=\min\limits_{1\leq i\leq N} k_i$ was proposed for HyperMap\cite{HyperMap}. We did not embed the American College Football network with $m=\min\limits_{1\leq i\leq N} k_i$, because this would have resulted in a negative $L=\left< k\right>/2-m$ value which was originally not allowed in HyperMap. For the Cambrian food web (in which the links are actually directed) the type of the node degree determining the radial order of the nodes was considered to be an additional parameter of the embedding: in the fourth and the fifth parameter setting procedures the simultaneous optimisation of $m$, $\beta$ and $T$ was carried out using such a radial node order that resulted in outwards decreasing in-degree and out-degree, respectively. All the parameter settings tested for ncMCE, its angular optimisation and HyperMap are given in Table \ref{table_realParameters}. In the case of Mercator, we used its own parameter estimation process included in the algorithm\cite{Boguna_embedding_2019}.

\begin{table}[!ht]
\centering
\caption{{\bf Properties of the studied real networks and the embedding parameters used for ncMCE, its angular optimisation and HyperMap.} The following observed network characteristics are listed: the number of nodes $N$, the average degree $\left< k\right>$, the minimal degree $\min\limits_{1\leq i\leq N} k_i$ and average clustering coefficient $\left< c\right>$. For each network at least two different procedures were carried out for setting the embedding parameters $m$, $L$, $\beta$ and $T$. The main idea behind our parameter estimation approach was to minimise the logarithmic loss $LL$ of an embedding obtained with the ncMCE method. Note that $\zeta$ was always set to $1$, %which corresponds to the standard practice in the literature, 
and $L$ was actually not considered to be a free parameter, instead it was always calculated  as $L=\left< k\right>/2-m$.}
\begin{tabular}{+c+c|c|c|c|c+}
\thickhline
& P. A. Collab. & Books about & Am. College & Cambr. food web & Prot. int. netw. \\ & SD-reconstr. &  US politics & Football & from the B. S. & from PDZBase \\ \thickhline

\rowcolor{lightgray} \multicolumn{6}{+c+}{\bf Measured network characteristics} \\ \hline %usage: \multicolumn{numberOfCols}{align}{text}, where align=l/c/r
\rowcolor{lightgray} $N$ & $38$ & $105$ & $115$ & $142$ & $161$\\ \hline
\rowcolor{lightgray} $\left< k\right>$ & $5.368$ & $8.4$ & $10.661$ & $10.761$ & $2.596$\\ \hline
\rowcolor{lightgray} $\min\limits_{1\leq i\leq N} k_i$ & $1$ & $2$ & $7$ & $1$ & $1$\\ \hline
\rowcolor{lightgray} $\left< c\right>$ & $0.806$ & $0.488$ & $0.403$ & $0.205$ & $0.007$\\ \thickhline

\rowcolor{white} \multicolumn{6}{+c+}{\bf Optimal embedding parameters according to a gradient descent}\\
\rowcolor{white}\multicolumn{6}{+c+}{\bf in the $m-\beta-T$ space with $\left< k\right>/2<m$ allowed} \\ \hline
\rowcolor{white} $m$ & $3.067$ & $5.437$ & $10.965$ & $6.233$ & $1.524$\\ \hline
\rowcolor{white} $L$ & $-0.383$ & $-1.237$ & $-5.635$ & $-0.853$ & $-0.226$\\ \hline
\rowcolor{white} $\beta$ & $0.561$ & $0.526$ & $0.1$ & $0.979$ & $0.591$\\ \hline
\rowcolor{white} $T$ & $0.347$ & $0.624$ & $0.577$ & $0.804$ & $0.536$\\ \thickhline

\rowcolor{lightgray} \multicolumn{6}{+c+}{\bf Optimal embedding parameters according to a gradient descent} \\
\rowcolor{lightgray} \multicolumn{6}{+c+}{\bf in the $\beta-T$ space using $m=\left< k\right>/2$ and $L=0$} \\ \hline
\rowcolor{lightgray} $\beta$ & $0.618$ & $0.579$ & $0.304$ & $0.888$ & $0.634$\\ \hline
\rowcolor{lightgray} $T$ & $0.379$ & $0.590$ & $0.588$ & $0.747$ & $0.519$\\ \thickhline

\rowcolor{white} \multicolumn{6}{+c+}{\bf Optimal embedding parameters according to a gradient descent} \\
\rowcolor{white} \multicolumn{6}{+c+}{\bf in the $\beta-T$ spaceusing $m=\min\limits_{1\leq i\leq N} k_i$ and $0<L=\left< k\right>/2-m$} \\ \hline
\rowcolor{white} $\beta$ & $0.939$ & $0.762$ & $-$ & $0.99$ & $0.688$\\ \hline
\rowcolor{white} $T$ & $0.504$ & $0.621$ & $-$ & $0.664$ & $0.543$\\ \thickhline

\rowcolor{lightgray} \multicolumn{6}{+c+}{\bf Optimal embedding parameters according to a gradient descent} \\
\rowcolor{lightgray} \multicolumn{6}{+c+}{\bf in the $m-\beta-T$ space with $\left< k\right>/2<m$ allowed}  \\
\rowcolor{lightgray} \multicolumn{6}{+c+}{\bf and a radial order according to the in-degree $k_{\mathrm{in}}$} \\ \hline
\rowcolor{lightgray} $m$ & $-$ & $-$ & $-$ & $9.976$ & $-$\\ \hline
\rowcolor{lightgray} $L$ & $-$ & $-$ & $-$ & $-4.596$ & $-$\\ \hline
\rowcolor{lightgray} $\beta$ & $-$ & $-$ & $-$ & $0.681$ & $-$\\ \hline
\rowcolor{lightgray} $T$ & $-$ & $-$ & $-$ & $0.853$ & $-$\\ \thickhline

\rowcolor{white} \multicolumn{6}{+c+}{\bf Optimal embedding parameters according to a gradient descent}\\
\rowcolor{white} \multicolumn{6}{+c+}{\bf in the $m-\beta-T$ space with $\left< k\right>/2<m$ allowed} \\
\rowcolor{white} \multicolumn{6}{+c+}{\bf and a radial order according to the out-degree $k_{\mathrm{out}}$} \\ \hline
\rowcolor{white} $m$ & $-$ & $-$ & $-$ & $8.642$ & $-$\\ \hline
\rowcolor{white} $L$ & $-$ & $-$ & $-$ & $-3.262$ & $-$\\ \hline
\rowcolor{white} $\beta$ & $-$ & $-$ & $-$ & $0.740$ & $-$\\ \hline
\rowcolor{white} $T$ & $-$ & $-$ & $-$ & $0.837$ & $-$\\ \thickhline
\end{tabular}
%\begin{flushleft} Table notes???
%\end{flushleft}
\label{table_realParameters}
\end{table}

\subsection{Embedding quality at different parameter settings}

\begin{figure}
    \centering
    \includegraphics[width=\textwidth]{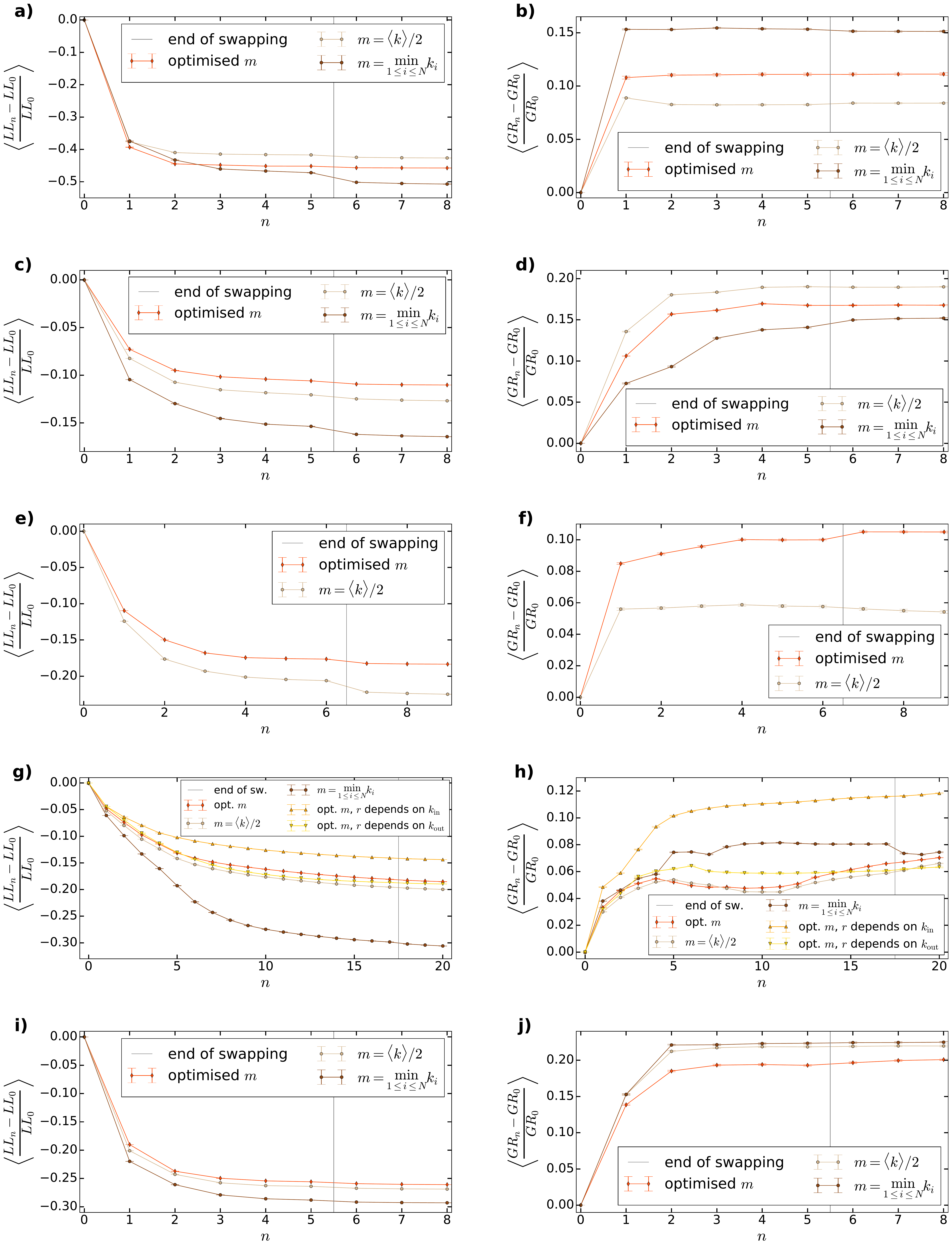}
    \caption{{\bf The convergence of the logarithmic loss $LL$ (left) and the greedy routing score $GR$ (right) over the subsequent rounds of iterations during the angular optimisation of the embeddings resulted from ncMCE for the studied real networks.} The curves of different colours show the relative improvement in the embedding quality for different settings of the embedding parameters: either $m$ was optimised simultaneously with $\beta$ and $T$ via a gradient descent, or only $\beta$ and $T$ were optimised based on the logarithmic loss and $m$ was set to the half of average degree or the smallest occurring degree. In the case of the Cambrian food web (panels g) and h)) two further parameter adjusting procedures were also carried out, where the parameter $m$ was optimised simultaneously with $\beta$ and $T$ for such node arrangements where the radial coordinates were assigned in the order of the in- or the out-degrees instead of the total degrees of the nodes. Each data point corresponds to an average over $2500$ trials of embedding with the given $m$, $\beta$ and $T$ parameters, and the bars indicate the $95\%$ confidence intervals. The total number of optimisation rounds was set for each network individually to a value at which the logarithmic loss seemed to settle to a more or less constant value with all the examined parameter settings. Panels a) and b): The twelfth layer of the Pierre Auger collaboration network. Panels  c) and d): Network between books about US politics. Panels e) and f): American College Football network. Panels g) and h): Cambrian food web from the Burgess Shale. Panels i) and j): Protein interaction network from the PDZBase database.}
    \label{fig:ncMCEimpr_real}
\end{figure}

In the following, we show detailed results regarding the behaviour of the quality scores as a function of the embedding parameters. As it can be seen in Fig. \ref{fig:ncMCEimpr_real}, both the logarithmic loss $LL$ and the greedy routing score $GR$ reach a more or less steady value for all the studied real networks and parameter settings within a reasonable number of rounds of angular optimisation of the node arrangement obtained with ncMCE, just like in the case of synthetic networks.

\begin{figure}
    \centering
    \includegraphics[width=\textwidth]{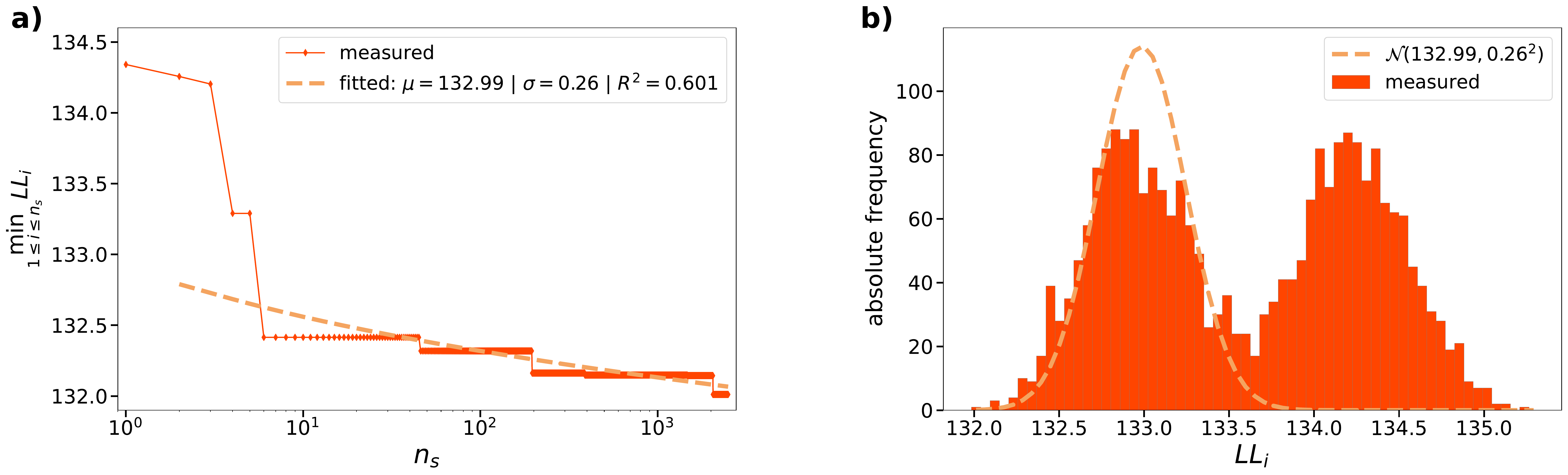}
    \caption{{\bf Example for fitting to the achieved best logarithmic loss as a function of the number of embedding trials for a real network where the distribution of the obtained quality scores among the repetitions of the embedding seems to be bi-modal.} The here considered $LL$ values resulted from the repeated embedding of the Pierre Auger collaboration network using the original ncMCE approach with the parameters $m$, $\beta$ and $T$ obtained from our parameter optimisation procedure described in a previous section. a) The achieved best $LL$ values as a function of the number of samples $n_s$, together with the fitted curve according to equations (\ref{eq:LL_extr}) and (\ref{eq:fittedFunctionsForMoreTrials}). b) The density function of the obtained $LL$ scores (shown by the histogram) together with a normal distribution (dashed line) having a mean and standard deviation corresponding to the fit shown in panel a), re-normalised according to the weight of the empirical $LL$ samples below $LL=133$.
%    Példa arra, hogy bár a valós hálók esetén az LL különböző beágyazási próbálkozások közötti eloszlása nem feltétlenül egy sima normális eloszlás (mint a szintetikus hálóknál), de az adott próbálkozási számig elért legjobb értékre illesztésnek itt is van értelme, sőt, ez több púp esetén automatikusan megtalálja az optimumhoz közelebbit, így kifejezetten érdemesebb a hisztogram helyett az adott próbálkozási számig elért legjobb értékek görbéjére illeszteni. Ugyanez igaz a GR-score-ra is... A példa a Pierre Auger collaboration network-nek a saját paramétermeghatározási módszerünkből (m, béta és T egyszerre optimalizálva) kapott paraméterek melletti különböző ncMCE-s beágyazásaira vonatkozó LL értékeket mutatja be. Az a) ábrán az elért legjobb LL és a rá illesztett függvény szerepel az $n_s$ próbálkozási szám függvényében. A b) ábrán a logloss különböző beágyazási próbálkozások közti eloszlása látható, amire rárajzoltam az illesztésnek megfelelő átlagú és szórású normális eloszlást, amit úgy normáltam, hogy a jobb púptól már egyértelműen leváló részén (132 és 133 között) az integrálja megegyezzen az ezen a tartományon mért hisztogram alatti területtel.}
}
    \label{fig:realTrialHist_moreThan1Peak}
\end{figure}

In the previous section we have shown that for synthetic networks the achievable improvement in the quality scores under the repetition of the embedding can be predicted for all the four studied embedding methods by fitting the formulae in equations (\ref{eq:LL_extr}-\ref{eq:fittedFunctionsForMoreTrials}) to the lowest achieved $LL$ or the highest achieved $GR$ as a function of the number of trials so far. In Fig. \ref{fig:realTrialHist_moreThan1Peak} we demonstrate that although in the case of real networks it is often not true anymore that the quality distribution among the repetitions of the embedding is a simple normal distribution, the corresponding formulae in equations (\ref{eq:LL_extr}-\ref{eq:fittedFunctionsForMoreTrials}) still can be used to fit the observed curves of the best quality scores achieved so far. The fitted coefficients $\mu$ and $\sigma$ correspond to the mean and the standard deviation of the peak found on that side of the quality distribution which yields the best results. Thus, it seems that fitting equations (\ref{eq:LL_extr}-\ref{eq:fittedFunctionsForMoreTrials}) automatically selects the part of the quality distribution that is relevant for estimating the quality improvement achievable by the repetition of the embedding. %[EZ ÚGY HANGZIK, MINTHA EGYÁLTALÁN NEM SZÁMÍTANA AZ ELOSZLÁS TÖBBI RÉSZE, PEDIG NYILVÁN A PRÓBÁK CSOMÓ RÉSZE NEM A KEDVEZŐBB CSÚCS TARTOMÁNYÁBA ESIK...]

\begin{figure}
    \centering
    \includegraphics[width=\textwidth]{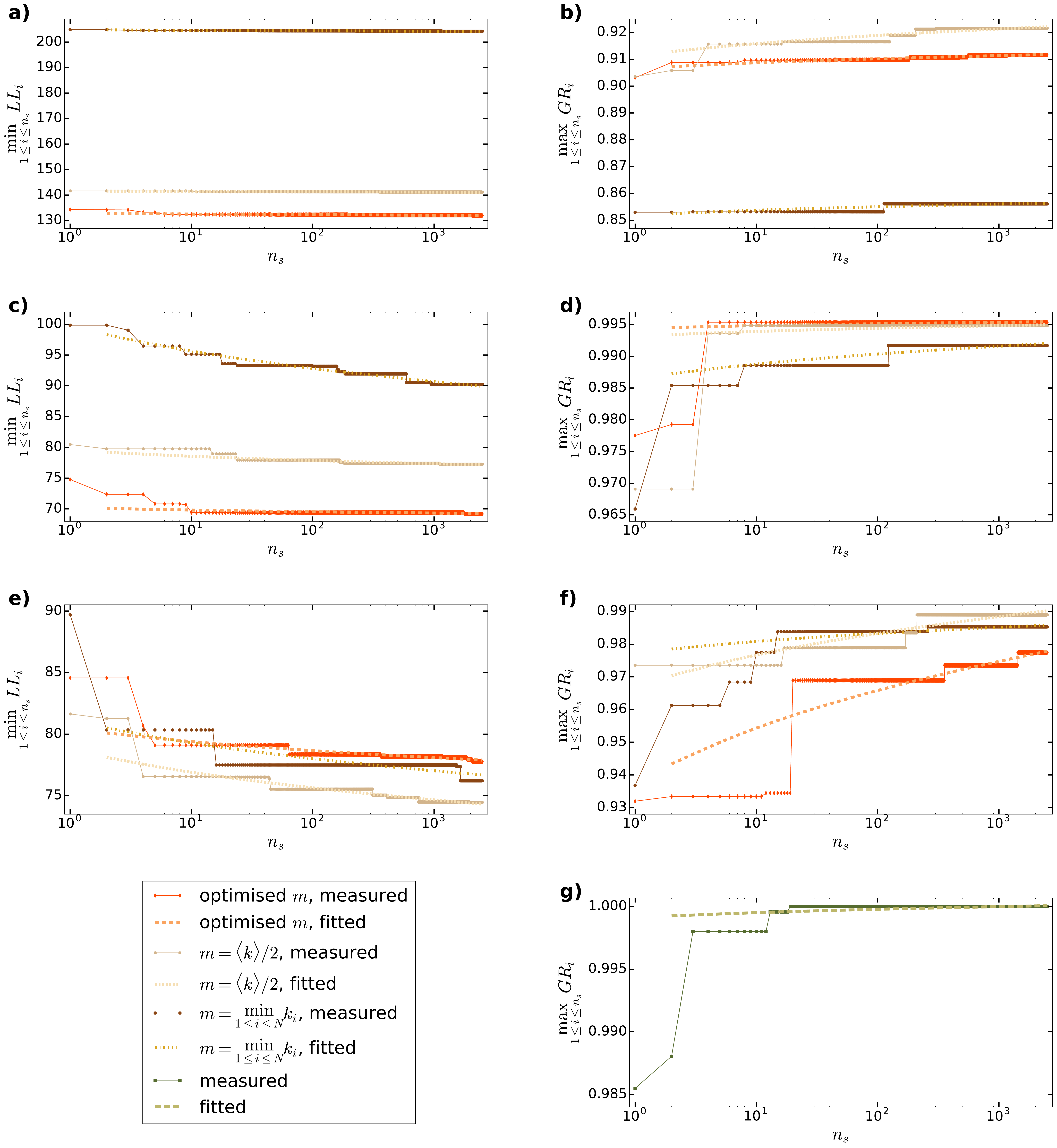}
    \caption{{\bf The achieved best logarithmic loss (left) and greedy routing score (right) as a function of the number of repetitions of the embedding in the case of the twelfth layer of the Pierre Auger collaboration network.} Each row of the figure presents the results regarding one of the studied embedding algorithms: panels a) and b) refer to ncMCE, panels c) and d) refer to ncMCE with angular optimisation, panels e) and f) refer to HyperMap and panel g) refers to Mercator. Excepting Mercator, all the embedding methods were tested using $3$ different parameter settings: either $m$ was optimised together with $\beta$ and $T$, or only $\beta$ and $T$ were optimised with $m$ fixed to the half of average degree or the smallest occurring degree. The dashed curves were obtained by fitting the corresponding formulae of equations (\ref{eq:LL_extr}-\ref{eq:fittedFunctionsForMoreTrials}) to the solid curves. The fitted coefficients and the quality of the fits are listed in Table \ref{table_moreTrialFitting_collab}.}
    \label{fig:imprWithTrials_collab}
\end{figure}

\begin{figure}
    \centering
    \includegraphics[width=\textwidth]{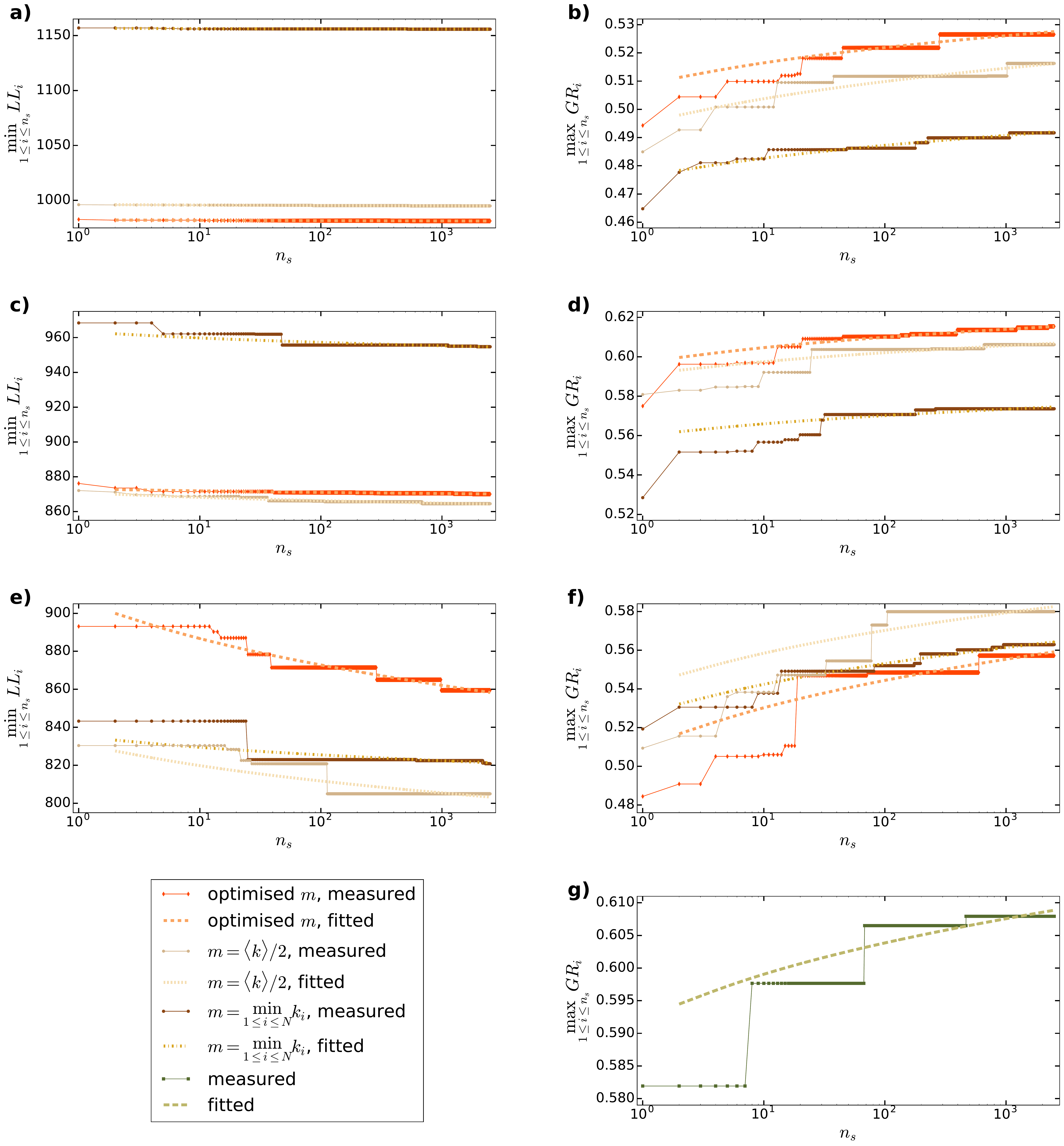}
    \caption{{\bf The achieved best logarithmic loss (left) and greedy routing score (right) as a function of the number of repetitions of the embedding in the case of the network between books about US politics.} Each row of the figure presents the results regarding one of the studied embedding algorithms: panels a) and b) refer to ncMCE, panels c) and d) refer to ncMCE with angular optimisation, panels e) and f) refer to HyperMap and panel g) refers to Mercator. Excepting Mercator, all the embedding methods were tested using $3$ different parameter settings: either $m$ was optimised together with $\beta$ and $T$, or only $\beta$ and $T$ were optimised with $m$ fixed to the half of average degree or the smallest occurring degree. The dashed curves were obtained by fitting the corresponding formulae in equations (\ref{eq:LL_extr}-\ref{eq:fittedFunctionsForMoreTrials}) to the solid curves. The fitted coefficients and the quality of the fits are listed in Table \ref{table_moreTrialFitting_polBooks}.}
    \label{fig:imprWithTrials_polBooks}
\end{figure}

\begin{figure}
    \centering
    \includegraphics[width=\textwidth]{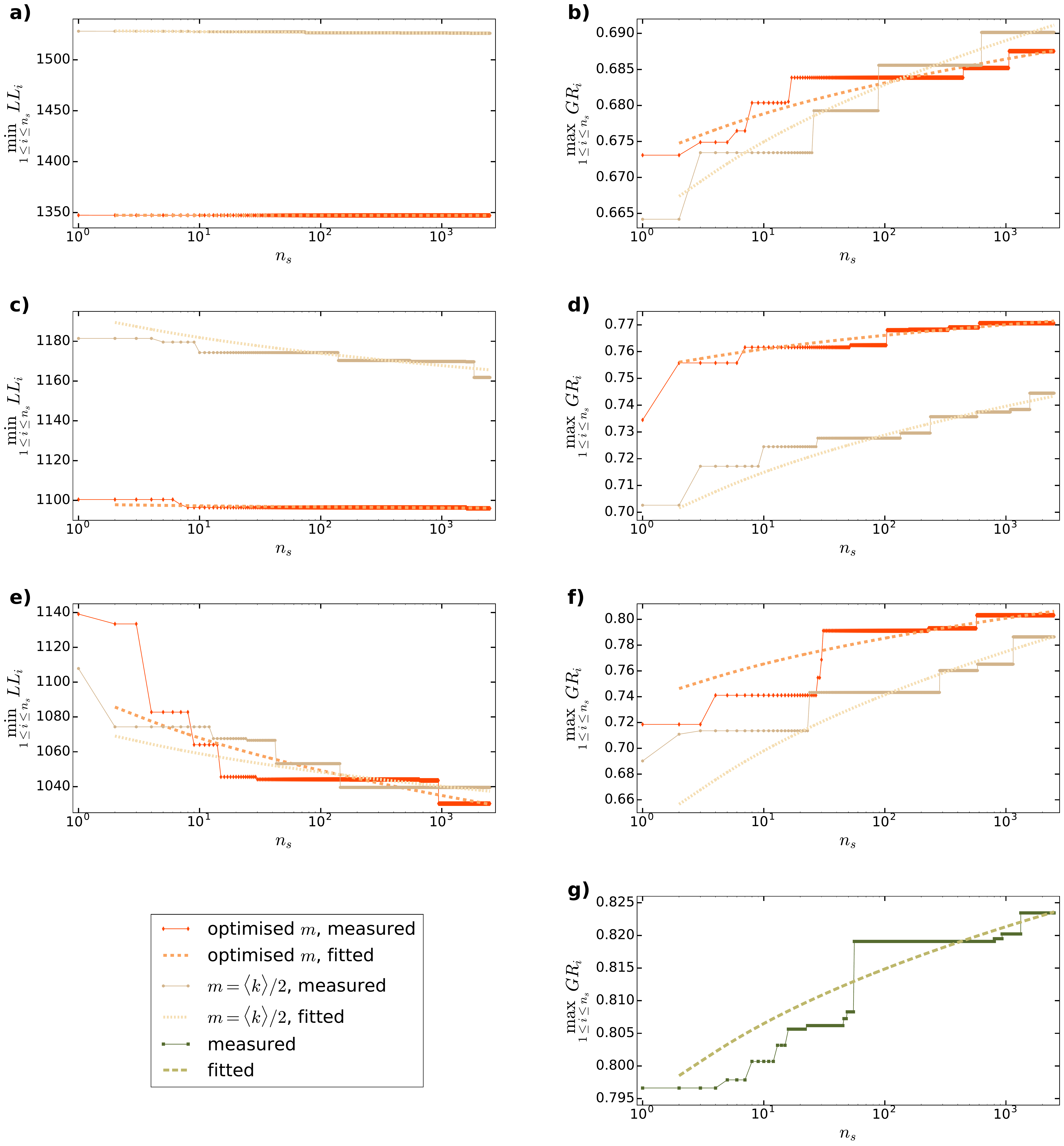}
    \caption{{\bf The achieved best logarithmic loss (left) and greedy routing score (right) as a function of the number of repetitions of the embedding in the case of the American College Football network.} Each row of the figure presents the results regarding one of the studied embedding algorithms: panels a) and b) refer to ncMCE, panels c) and d) refer to ncMCE with angular optimisation, panels e) and f) refer to HyperMap and panel g) refers to Mercator. Excepting Mercator, all the embedding methods were tested using $2$ different parameter settings: either $m$ was optimised together with $\beta$ and $T$, or only $\beta$ and $T$ were optimised with $m$ fixed to the half of average degree. The dashed curves were obtained by fitting the corresponding formulae of equations (\ref{eq:LL_extr}-\ref{eq:fittedFunctionsForMoreTrials}) to the solid curves. The fitted coefficients and the quality of the fits are listed in Table \ref{table_moreTrialFitting_football}.}
    \label{fig:imprWithTrials_football}
\end{figure}

\begin{figure}
    \centering
    \includegraphics[width=\textwidth]{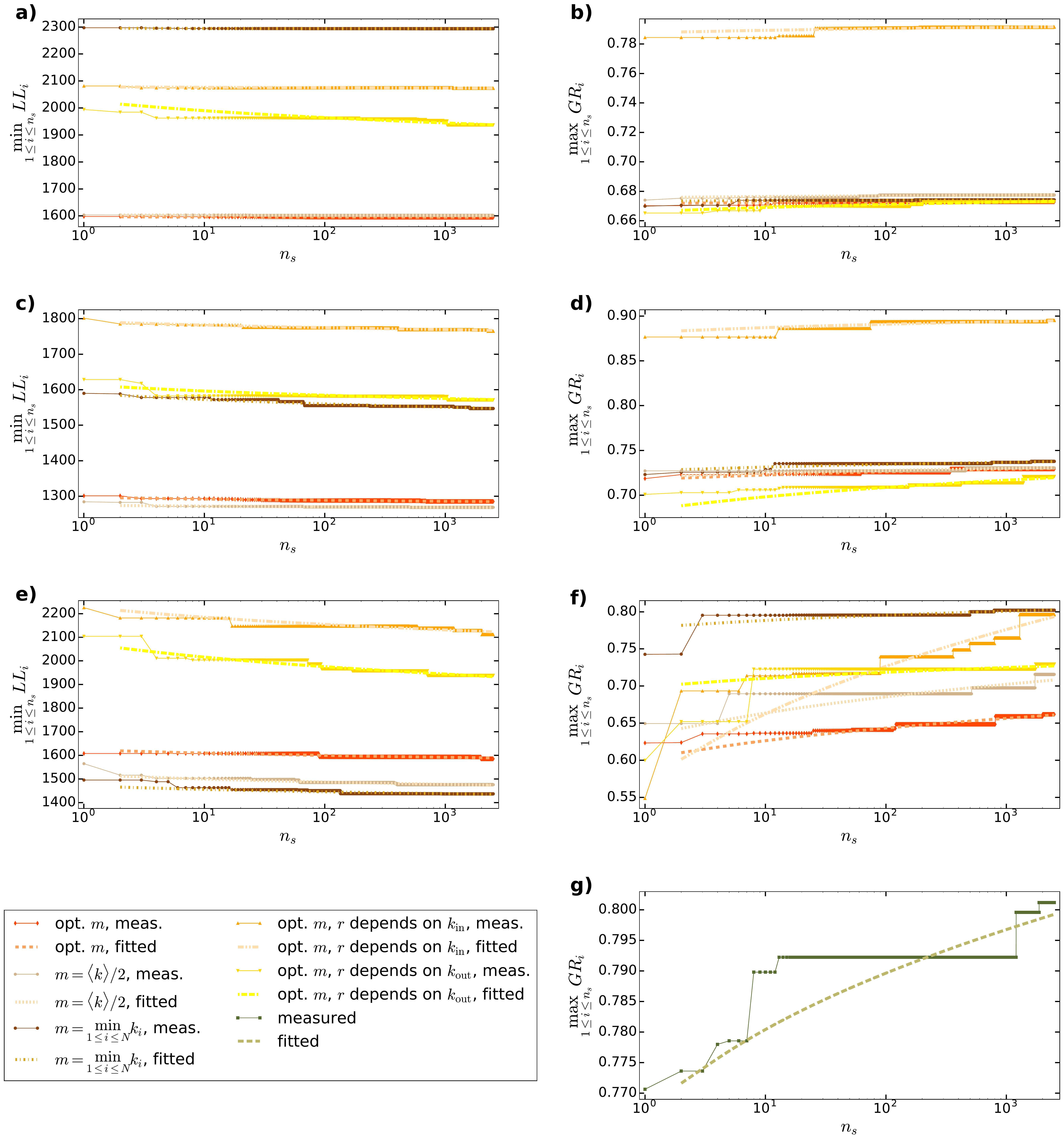}
    \caption{{\bf The achieved best logarithmic loss (left) and greedy routing score (right) as a function of the number of repetitions of the embedding in the case of the Cambrian food web from the Burgess Shale.} Each row of the figure presents the results regarding one of the studied embedding algorithms: panels a) and b) refer to ncMCE, panels c) and d) refer to ncMCE with angular optimisation, panels e) and f) refer to HyperMap and panel g) refers to Mercator. Excepting Mercator, all the embedding methods were tested using $5$ different parameter settings: either $m$ was optimised together with $\beta$ and $T$ using a radial order determined by the total, the in- or the out-degree of the nodes, or only $\beta$ and $T$ were optimised with $m$ fixed to the half of average degree or the smallest occurring degree. The dashed curves were obtained by fitting the corresponding formulae of equations (\ref{eq:LL_extr}-\ref{eq:fittedFunctionsForMoreTrials}) to the solid curves. The fitted coefficients and the quality of the fits are listed in Table \ref{table_moreTrialFitting_foodWeb}.}
    \label{fig:imprWithTrials_foodWeb}
\end{figure}

\begin{figure}
    \centering
    \includegraphics[width=\textwidth]{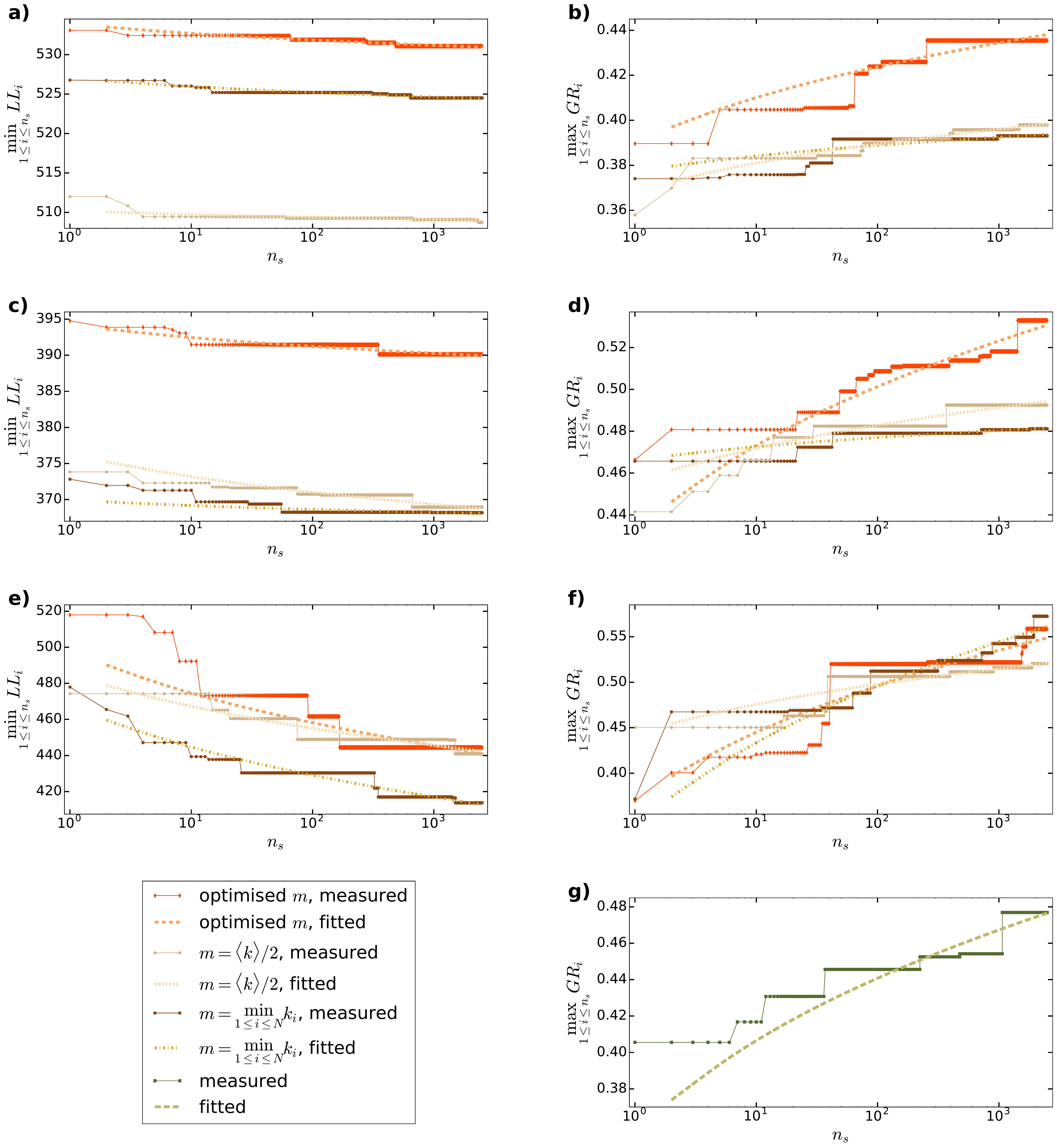}
    \caption{{\bf The achieved best logarithmic loss (left) and greedy routing score (right) as a function of the number of repetitions of the embedding in the case of the protein interaction network from the PDZBase database.} Each row of the figure presents the results regarding one of the studied embedding algorithms: panels a) and b) refer to ncMCE, panels c) and d) refer to ncMCE with angular optimisation, panels e) and f) refer to HyperMap and panel g) refers to Mercator. Excepting Mercator, all the embedding methods were tested using $3$ different parameter settings: either $m$ was optimised together with $\beta$ and $T$, or only $\beta$ and $T$ were optimised with $m$ fixed to the half of average degree or the smallest occurring degree. The dashed curves were obtained by fitting the corresponding formulae of equations (\ref{eq:LL_extr}-\ref{eq:fittedFunctionsForMoreTrials}) to the solid curves. The fitted coefficients and the quality of the fits are listed in Table \ref{table_moreTrialFitting_proteinInt}.}
    \label{fig:imprWithTrials_proteinInt}
\end{figure}

%Because of the variance in the embedding quality appearing with/during(?) the repetition of the embedding process
We embedded all the five real networks with each of the examined embedding methods $2500$ times. In the case of ncMCE, ncMCE with angular optimisation and HyperMap we set the embedding parameters only once for each network with each of the above described parameter estimating methods and during the repetition of the embedding only the radial order of the nodes having the same degree was randomly permuted again and again. Since Mercator has its own parameter estimating process and does not allow the setting of the radial order of the nodes, Mercator was simply re-run $2500$ times. The achieved best quality scores are plotted as a function of the number of repetitions $n_s$ in Figs. \ref{fig:imprWithTrials_collab}-\ref{fig:imprWithTrials_proteinInt}. We fitted to each curve according to equations (\ref{eq:LL_extr}-\ref{eq:fittedFunctionsForMoreTrials}), the fitted coefficients are listed in Tables \ref{table_moreTrialFitting_collab}-\ref{table_moreTrialFitting_proteinInt}. An interesting point to note related to these results is that the greedy routing score achieved for the Cambrian food web was way higher when the radial ordering of the nodes was dictated by the in-degree instead of the out-degree or the total degree for both the original ncMCE approach and its angularly optimised version.

\begin{table}[!ht]
\centering
\caption{{\bf The results of fitting the formulae in equations (\ref{eq:LL_extr}-\ref{eq:fittedFunctionsForMoreTrials}) to the curves plotted in Fig. \ref{fig:imprWithTrials_collab} showing the achieved best quality scores as a function of the number of embeddings carried out in the case of the twelfth layer of the Pierre Auger collaboration network.} The fitted coefficients are the mean ($\mu_{LL}$ and $\mu_{GR}$) and the standard deviation ($\sigma_{LL}$ and $\sigma_{GR}$) characterising the peak on that side of the quality score's observed distribution which corresponds to the best results. The quality of the fits is characterised by the coefficient of determination ($R_{LL}^2$ and $R_{GR}^2$), which is a sufficiently high value in most of the cases. For the three embedding methods that were tried out with various parametrisations, the results regarding the eventually (after $2500$ repetitions) best parameter setting are written in bold.}
\begin{tabular}{+l+c|c|c+c|c|c+}
\thickhline
\rowcolor{white} Embedding method & $\mu_{LL}$ & $\sigma_{LL}$ & $R^2_{LL}$ & $\mu_{GR}$ & $\sigma_{GR}$ & $R^2_{GR}$ \\ \thickhline
\rowcolor{lightgray} ncMCE, optimised $m$ & $\pmb{132.99}$ & $\pmb{0.261}$ & $\pmb{0.60}$ & $0.906$ & $0.0016$ & $0.87$ \\ \hline
\rowcolor{lightgray} ncMCE, $m=\left< k\right>/2$ and $L=0$ & $141.70$ & $0.157$ & $0.79$ & $\pmb{0.910}$ & $\pmb{0.0033}$ & $\pmb{0.67}$ \\ \hline
\rowcolor{lightgray} ncMCE, $m=\min\limits_{1\leq i\leq N} k_i$ and $0<L=\left< k\right>/2-m$ & $204.91$ & $0.202$ & $0.77$ & $0.851$ & $0.0014$ & $0.49$ \\ \hline
\rowcolor{white} ncMCE with ang. opt., optimised $m$ & $\pmb{70.29}$ & $\pmb{0.296}$ & $\pmb{0.34}$ & $\pmb{0.994}$ & $\pmb{0.0003}$ & $\pmb{0.05}$ \\ \hline
\rowcolor{white} ncMCE with ang. opt., $m=\left< k\right>/2$ and $L=0$ & $79.75$ & $0.738$ & $0.77$ & $0.993$ & $0.0006$ & $0.06$ \\ \hline
\rowcolor{white} ncMCE with ang. opt., $m=\min\limits_{1\leq i\leq N} k_i$  & $100.56$ & $3.020$ & $0.89$ & $0.986$ & $0.0017$ & $0.54$ \\ 
\rowcolor{white} and $0<L=\left< k\right>/2-m$ &  &  &  &  &  &  \\ \hline
\rowcolor{lightgray} HyperMap, optimised $m$ & $80.69$ & $0.795$ & $0.65$ & $0.934$ & $0.0124$ & $0.73$ \\ \hline
\rowcolor{lightgray} HyperMap, $m=\left< k\right>/2$ and $L=0$ & $\pmb{79.14}$ & $\pmb{1.369}$ & $\pmb{0.86}$ & $\pmb{0.965}$ & $\pmb{0.0071}$ & $\pmb{0.65}$ \\ \hline
\rowcolor{lightgray} HyperMap, $m=\min\limits_{1\leq i\leq N} k_i$ and $0<L=\left< k\right>/2-m$ & $81.54$ & $1.381$ & $0.43$ & $0.977$ & $0.0026$ & $0.40$ \\ \hline
\rowcolor{white} Mercator & $-$ & $-$ & $-$ & $0.999$ & $0.0003$ & $0.11$ \\ \thickhline
\end{tabular}
%\begin{flushleft} Table notes???
%\end{flushleft}
\label{table_moreTrialFitting_collab}
\end{table}

\begin{table}[!ht]
\centering
\caption{{\bf The results of fitting the formulae in equations (\ref{eq:LL_extr}-\ref{eq:fittedFunctionsForMoreTrials}) to the curves plotted in Fig. \ref{fig:imprWithTrials_polBooks} showing the achieved best quality scores as a function of the number of embeddings carried out in the case of the network between books about US politics.} The fitted coefficients are the mean ($\mu_{LL}$ and $\mu_{GR}$) and the standard deviation ($\sigma_{LL}$ and $\sigma_{GR}$) characterising the peak on that side of the quality score's observed distribution which corresponds to the best results. The quality of the fits is characterised by the coefficient of determination ($R_{LL}^2$ and $R_{GR}^2$), which is a sufficiently high value in most of the cases. For the three embedding methods that were tried out with various parametrisations, the results regarding the eventually (after $2500$ repetitions) best parameter setting are written in bold.}
\begin{tabular}{+l+c|c|c+c|c|c+}
\thickhline
\rowcolor{white} Embedding method & $\mu_{LL}$ & $\sigma_{LL}$ & $R^2_{LL}$ & $\mu_{GR}$ & $\sigma_{GR}$ & $R^2_{GR}$ \\ \thickhline
\rowcolor{lightgray} ncMCE, optimised $m$ & $\pmb{982.38}$ & $\pmb{0.321}$ & $\pmb{0.78}$ & $\pmb{0.51}$ & $\pmb{0.006}$ & $\pmb{0.72}$ \\ \hline
\rowcolor{lightgray} ncMCE, $m=\left< k\right>/2$ and $L=0$ & $996.44$ & $0.425$ & $0.89$ & $0.49$ & $0.007$ & $0.67$ \\ \hline
\rowcolor{lightgray} ncMCE, $m=\min\limits_{1\leq i\leq N} k_i$ and $0<L=\left< k\right>/2-m$ & $1156.67$ & $0.264$ & $0.72$ & $0.48$ & $0.005$ & $0.88$ \\ \hline
\rowcolor{white} ncMCE with ang. opt., optimised $m$ & $873.37$ & $0.917$ & $0.83$ & $\pmb{0.60}$ & $\pmb{0.006}$ & $\pmb{0.88}$ \\ \hline
\rowcolor{white} ncMCE with ang. opt., $m=\left< k\right>/2$ and $L=0$ & $\pmb{871.51}$ & $\pmb{2.045}$ & $\pmb{0.82}$ & $0.59$ & $0.005$ & $0.67$ \\ \hline
\rowcolor{white} ncMCE with ang. opt., $m=\min\limits_{1\leq i\leq N} k_i$ & $964.21$ & $2.734$ & $0.62$ & $0.56$ & $0.005$ & $0.54$ \\ 
\rowcolor{white} and $0<L=\left< k\right>/2-m$ & &  &  &  & &  \\ \hline
\rowcolor{lightgray} HyperMap, optimised $m$ & $911.24$ & $15.031$ & $0.89$ & $0.51$ & $0.015$ & $0.71$ \\ \hline
\rowcolor{lightgray} HyperMap, $m=\left< k\right>/2$ and $L=0$ & $\pmb{834.06}$ & $\pmb{8.748}$ & $\pmb{0.53}$ & $\pmb{0.54}$ & $\pmb{0.013}$ & $\pmb{0.51}$ \\ \hline
\rowcolor{lightgray} HyperMap, $m=\min\limits_{1\leq i\leq N} k_i$ and $0<L=\left< k\right>/2-m$ & $836.44$ & $4.295$ & $0.40$ & $0.52$ & $0.012$ & $0.91$ \\ \hline
\rowcolor{white} Mercator & $-$ & $-$ & $-$ & $0.59$ & $0.005$ & $0.61$ \\ \thickhline
\end{tabular}
%\begin{flushleft} Table notes???
%\end{flushleft}
\label{table_moreTrialFitting_polBooks}
\end{table}

\begin{table}[!ht]
\centering
\caption{{\bf The results of fitting the formulae in equations (\ref{eq:LL_extr}-\ref{eq:fittedFunctionsForMoreTrials}) to the curves plotted in Fig. \ref{fig:imprWithTrials_football} showing the achieved best quality scores as a function of the number of embeddings carried out in the case of the American College Football network}. The fitted coefficients are the mean ($\mu_{LL}$ and $\mu_{GR}$) and the standard deviation ($\sigma_{LL}$ and $\sigma_{GR}$) characterising the peak on that side of the quality score's observed distribution which corresponds to the best results. The quality of the fits is characterised by the coefficient of determination ($R_{LL}^2$ and $R_{GR}^2$), which is a sufficiently high value in most of the cases. For the three embedding methods that were tried out with various parametrisations, the results regarding the eventually (after $2500$ repetitions) best parameter setting are written in bold.}
\begin{tabular}{+l+c|c|c+c|c|c+}
\thickhline
\rowcolor{white} Embedding method & $\mu_{LL}$ & $\sigma_{LL}$ & $R^2_{LL}$ & $\mu_{GR}$ & $\sigma_{GR}$ & $R^2_{GR}$ \\ \thickhline
\rowcolor{lightgray} ncMCE, optimised $m$ & $\pmb{1347.5}$ & $\pmb{0.10}$ & $\pmb{0.88}$ & $0.67$ & $0.005$ & $0.78$ \\ \hline
\rowcolor{lightgray} ncMCE, $m=\left< k\right>/2$ and $L=0$ & $1529.0$ & $0.85$ & $0.86$ & $\pmb{0.66}$ & $\pmb{0.009}$ & $\pmb{0.84}$ \\ \hline
\rowcolor{white} ncMCE with ang. opt., optimised $m$ & $\pmb{1098.3}$ & $\pmb{0.61}$ & $\pmb{0.40}$ & $\pmb{0.75}$ & $\pmb{0.006}$ & $\pmb{0.82}$ \\ \hline
\rowcolor{white} ncMCE with ang. opt., $m=\left< k\right>/2$ and $L=0$ & $1195.9$ & $8.60$ & $0.46$ & $0.69$ & $0.015$ & $0.83$ \\ \hline
\rowcolor{lightgray} HyperMap, optimised $m$ & $\pmb{1100.8}$ & $\pmb{20.14}$ & $\pmb{0.66}$ & $\pmb{0.73}$ & $\pmb{0.022}$ & $\pmb{0.72}$ \\ \hline
\rowcolor{lightgray} HyperMap, $m=\left< k\right>/2$ and $L=0$ & $1077.6$ & $11.44$ & $0.59$ & $0.62$ & $0.047$ & $0.82$ \\ \hline
\rowcolor{white} Mercator & $-$ & $-$ & $-$ & $0.79$ & $0.009$ & $0.77$ \\ \thickhline
\end{tabular}
%\begin{flushleft} Table notes???
%\end{flushleft}
\label{table_moreTrialFitting_football}
\end{table}

\begin{table}[!ht]
\centering
\caption{{\bf The results of fitting the formulae in equations (\ref{eq:LL_extr}-\ref{eq:fittedFunctionsForMoreTrials}) to the curves plotted in Fig. \ref{fig:imprWithTrials_foodWeb} showing the achieved best quality scores as a function of the number of embeddings carried out in the case of the Cambrian food web from the Burgess Shale.} The fitted coefficients are the mean ($\mu_{LL}$ and $\mu_{GR}$) and the standard deviation ($\sigma_{LL}$ and $\sigma_{GR}$) characterising the peak on that side of the quality score's observed distribution which corresponds to the best results. The quality of the fits is characterised by the coefficient of determination ($R_{LL}^2$ and $R_{GR}^2$), which is a sufficiently high value in most of the cases. For the three embedding methods that were tried out with various parametrisations, the results regarding the eventually (after $2500$ repetitions) best parameter setting are written in bold.}
\begin{tabular}{+l+c|c|c+c|c|c+}
\thickhline
\rowcolor{white} Embedding method & $\mu_{LL}$ & $\sigma_{LL}$ & $R^2_{LL}$ & $\mu_{GR}$ & $\sigma_{GR}$ & $R^2_{GR}$ \\ \thickhline
\rowcolor{lightgray} ncMCE, optimised $m$ & $\pmb{1596.8}$ & $\pmb{0.59}$ & $\pmb{0.64}$ & $0.672$ & $0.0004$ & $0.43$ \\ \hline
\rowcolor{lightgray} ncMCE, $m=\left< k\right>/2$ and $L=0$ & $1603.8$ & $0.63$ & $0.79$ & $0.676$ & $0.0006$ & $0.52$ \\ \hline
\rowcolor{lightgray} ncMCE, $m=\min\limits_{1\leq i\leq N} k_i$ & $2295.4$ & $0.50$ & $0.76$ & $0.673$ & $0.0003$ & $0.30$ \\ 
\rowcolor{lightgray} and $0<L=\left< k\right>/2-m$ &  &  &  &  &  & \\ \hline
\rowcolor{lightgray} ncMCE, optimised $m$, & $2079.0$ & $1.58$ & $0.48$ & $\pmb{0.787}$ & $\pmb{0.0012}$ & $\pmb{0.37}$ \\ 
\rowcolor{lightgray} radial order according to $k_{\mathrm{in}}$ &  &  &  &  & &  \\ \hline
\rowcolor{lightgray} ncMCE, optimised $m$,  & $2034.9$ & $27.64$ & $0.67$ & $0.665$ & $0.0022$ & $0.65$ \\ 
\rowcolor{lightgray} radial order according to $k_{\mathrm{out}}$ &  &  &  &  &  & \\ \hline
\rowcolor{white} ncMCE with ang. opt., optimised $m$ & $1298.3$ & $3.80$ & $0.82$ & $0.716$ & $0.0039$ & $0.71$ \\ \hline
\rowcolor{white} ncMCE with ang. opt., $m=\left< k\right>/2$ and $L=0$ & $\pmb{1275.3}$ & $\pmb{1.93}$ & $\pmb{0.71}$ & $0.719$ & $0.0033$ & $0.63$ \\ \hline
\rowcolor{white} ncMCE with ang. opt., $m=\min\limits_{1\leq i\leq N} k_i$  & $1593.7$ & $13.14$ & $0.82$ & $0.726$ & $0.0033$ & $0.70$ \\
\rowcolor{white} and $0<L=\left< k\right>/2-m$ &  & & &  &  &  \\ \hline
\rowcolor{white} ncMCE with ang. opt., optimised $m$, & $1794.7$ & $7.81$ & $0.80$ & $\pmb{0.880}$ & $\pmb{0.0041}$ & $\pmb{0.53}$ \\ 
\rowcolor{white} radial order according to $k_{\mathrm{in}}$ & & & & &  &  \\ \hline
\rowcolor{white} ncMCE with ang. opt., optimised $m$, & $1617.2$ & $12.85$ & $0.57$ & $0.680$ & $0.0113$ & $0.68$ \\ 
\rowcolor{white} radial order according to $k_{\mathrm{out}}$ & &  &  &  &  &  \\ \hline
\rowcolor{lightgray} HyperMap, optimised $m$ & $1626.9$ & $10.74$ & $0.55$ & $0.596$ & $0.0183$ & $0.80$ \\ \hline
\rowcolor{lightgray} HyperMap, $m=\left< k\right>/2$ and $L=0$ & $1522.1$ & $13.82$ & $0.79$ & $0.625$ & $0.0236$ & $0.51$ \\ \hline
\rowcolor{lightgray} HyperMap, $m=\min\limits_{1\leq i\leq N} k_i$  & $\pmb{1473.5}$ & $\pmb{10.99}$ & $\pmb{0.62}$ & $\pmb{0.776}$ & $\pmb{0.0076}$ & $\pmb{0.68}$ \\ 
\rowcolor{lightgray} and $0<L=\left< k\right>/2-m$ &  &  & &  &  & \\ \hline
\rowcolor{lightgray} HyperMap, optimised $m$,  & $2238.2$ & $32.71$ & $0.63$ & $0.549$ & $0.0693$ & $0.79$ \\
\rowcolor{lightgray} radial order according to $k_{\mathrm{in}}$ &  &  &  &  & & \\ \hline
\rowcolor{lightgray} HyperMap, optimised $m$,  & $2087.4$ & $43.79$ & $0.86$ & $0.696$ & $0.0089$ & $0.33$ \\ 
\rowcolor{lightgray} radial order according to $k_{\mathrm{out}}$ &  & & &  &  & \\ \hline
\rowcolor{white} Mercator & $-$ & $-$ & $-$ & $0.764$ & $0.0100$ & $0.53$ \\ \thickhline
\end{tabular}
%\begin{flushleft} Table notes???
%\end{flushleft}
\label{table_moreTrialFitting_foodWeb}
\end{table}

\begin{table}[!ht]
\centering
\caption{{\bf The results of fitting the formulae in equations (\ref{eq:LL_extr}-\ref{eq:fittedFunctionsForMoreTrials}) to the curves plotted in Fig. \ref{fig:imprWithTrials_proteinInt} showing the achieved best quality scores as a function of the number of embeddings carried out in the case of the protein interaction network from the PDZBase database.} The fitted coefficients are the mean ($\mu_{LL}$ and $\mu_{GR}$) and the standard deviation ($\sigma_{LL}$ and $\sigma_{GR}$) characterising the peak on that side of the quality score's observed distribution which corresponds to the best results. The quality of the fits is characterised by the coefficient of determination ($R_{LL}^2$ and $R_{GR}^2$), which is a sufficiently high value in most of the cases. For the three embedding methods that were tried out with various parametrisations, the results regarding the eventually (after $2500$ repetitions) best parameter setting are written in bold.}
\begin{tabular}{+l+c|c|c+c|c|c+}
\thickhline
\rowcolor{white} Embedding method & $\mu_{LL}$ & $\sigma_{LL}$ & $R^2_{LL}$ & $\mu_{GR}$ & $\sigma_{GR}$ & $R^2_{GR}$ \\ \thickhline
\rowcolor{lightgray} ncMCE, optimised $m$ & $534.17$ & $0.916$ & $0.82$ & $\pmb{0.39}$ & $\pmb{0.015}$ & $\pmb{0.69}$ \\ \hline
\rowcolor{lightgray} ncMCE, $m=\left< k\right>/2$ and $L=0$ & $\pmb{510.35}$ & $\pmb{0.378}$ & $\pmb{0.57}$ & $0.37$ & $0.009$ & $0.91$ \\ \hline
\rowcolor{lightgray} ncMCE, $m=\min\limits_{1\leq i\leq N} k_i$ and $0<L=\left< k\right>/2-m$ & $527.28$ & $0.812$ & $0.81$ & $0.38$ & $0.005$ & $0.59$ \\ \hline
\rowcolor{white} ncMCE with ang. opt., optimised $m$ & $394.64$ & $1.334$ & $0.69$ & $\pmb{0.42}$ & $\pmb{0.031}$ & $\pmb{0.78}$ \\ \hline
\rowcolor{white} ncMCE with ang. opt., $m=\left< k\right>/2$ and $L=0$ & $376.99$ & $2.335$ & $0.76$ & $0.45$ & $0.012$ & $0.74$ \\ \hline
\rowcolor{white} ncMCE with ang. opt., $m=\min\limits_{1\leq i\leq N} k_i$  & $\pmb{370.11}$ & $\pmb{0.574}$ & $\pmb{0.46}$ & $0.47$ & $0.005$ & $0.72$ \\ \hline
\rowcolor{white} and $0<L=\left< k\right>/2-m$ &  &  & &  &  &  \\ \hline
\rowcolor{lightgray} HyperMap, optimised $m$ & $503.44$ & $17.709$ & $0.62$ & $0.36$ & $0.055$ & $0.60$ \\ \hline
\rowcolor{lightgray} HyperMap, $m=\left< k\right>/2$ and $L=0$ & $488.71$ & $13.157$ & $0.67$ & $0.44$ & $0.024$ & $0.79$ \\ \hline
\rowcolor{lightgray} HyperMap, $m=\min\limits_{1\leq i\leq N} k_i$ and $0<L=\left< k\right>/2-m$ & $\pmb{472.22}$ & $\pmb{16.882}$ & $\pmb{0.84}$ & $\pmb{0.32}$ & $\pmb{0.068}$ & $\pmb{0.83}$ \\ \hline
\rowcolor{white} Mercator & $-$ & $-$ & $-$ & $0.35$ & $0.037$ & $0.74$ \\ \thickhline
\end{tabular}
%\begin{flushleft} Table notes???
%\end{flushleft}
\label{table_moreTrialFitting_proteinInt}
\end{table}

\subsection*{Hyperbolic layouts for real networks}

\begin{figure}
    \centering
    \includegraphics[width=\textwidth]{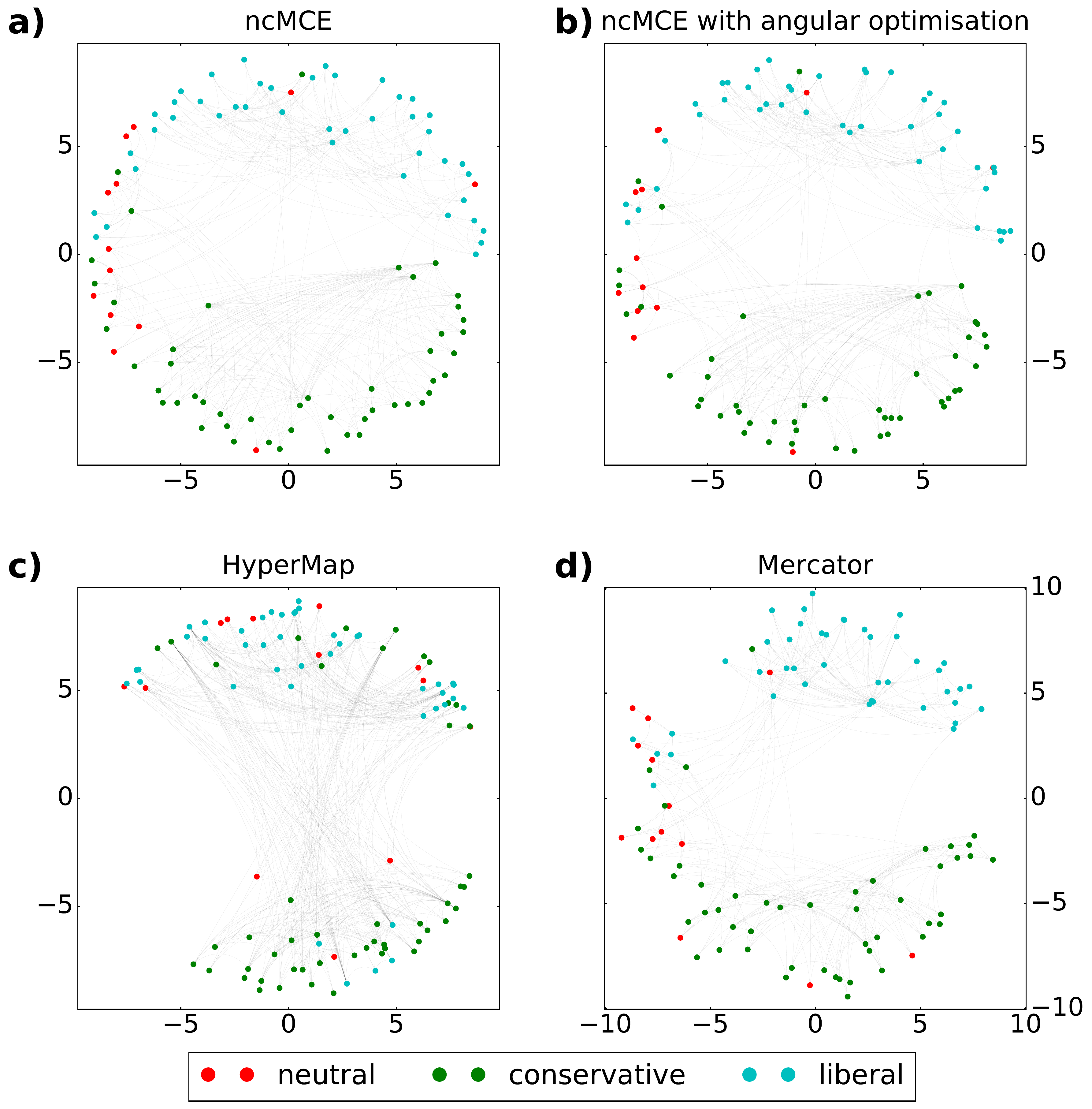}
    \caption{{\bf The layouts of the network between books about US politics on the native hyperbolic disk that reached the highest greedy routing scores.} a) The layout based on the coordinates resulted from the original ncMCE method. b) The layout according to the coordinates obtained with our approach, optimising the results of ncMCE. c) The hyperbolic layout obtained with HyperMap. d) The embedding according to Mercator.}
    \label{fig:polBooksPoinc}
\end{figure}

\begin{figure}
    \centering
    \includegraphics[width=\textwidth]{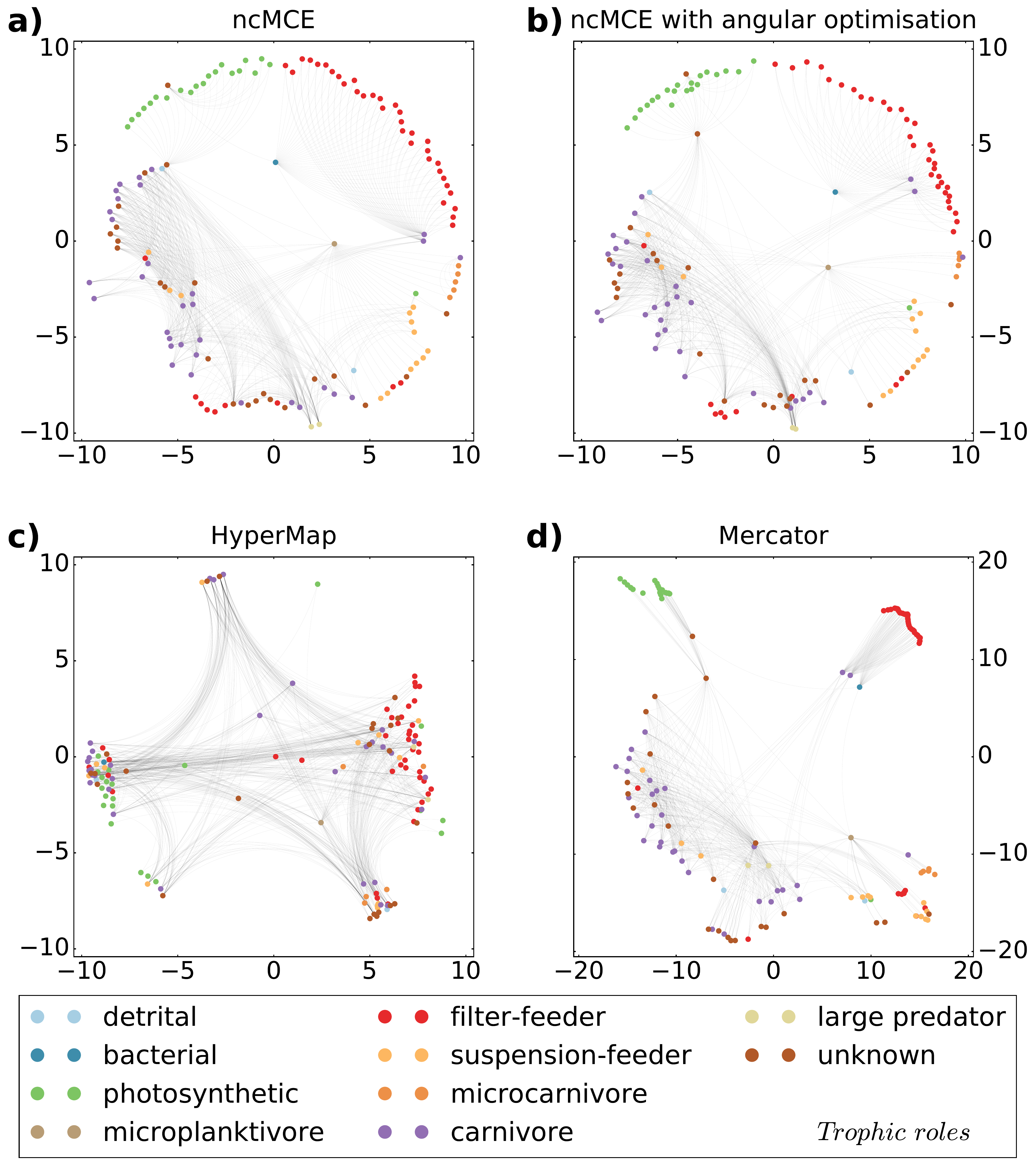}
    \caption{{\bf The layouts of the Cambrian food web on the native hyperbolic disk that reached the highest greedy routing scores.} a) The layout based on the coordinates resulted from the original ncMCE method. b) The layout according to the coordinates obtained with our approach, optimising the results of ncMCE. c) The hyperbolic layout obtained with HyperMap. d) The embedding according to Mercator.}
    \label{fig:foodWebPoinc}
\end{figure}

In addition to the hyperbolic layouts of the American College Football web presented in the main article, here we show layouts on the native representation of the hyperbolic plane also for the network between books about US politics and the Cambrian food web from the Burgess Shale. Similarly to the American College Football network, the nodes in these two further networks form communities. In the case of the network between books about US politics, shown in Fig. \ref{fig:polBooksPoinc}, the communities are based on political orientation, whereas for the Cambrian food web, depicted in Fig. \ref{fig:foodWebPoinc}, the nodes can be sorted into different trophic roles. For both networks, we have chosen the layout corresponding to the achieved best greedy routing score for each embedding method. 

%Az egyetlen, ami kifejezetten a radiális sorrend variálásáról szól az az, amikor azt nézzük, hogy segít-e a foodweben az irányítottság figyelembevétele, a többi esetben inkább a beágyazás többszöri elvégzéséről lenne célszerű beszélni, mivel ez értelmezhető a Mercatorra is.
%A Mercatornak csak egy irányítatlan éllista adható meg bemenetként. Bár a szogkoordináták keresésénél már mi is irányítatlannak tekintjük az éleket (ncMCE, szögoptimalizáló irányítatlan éllistával dolgozik), a radiális sorrend felállításakor (degree helyett in-/out-degree szerinti sorrend alkalmazásával) mi figyelembe tudjuk venni az irányítottságot.

%Célszerű lehet kihasználni a háló irányítottságát (ezt eddig mindenhol figyelmen kívül hagyták) a radiális sorrend felállításakor, mivel ez a foodweb-es próbálkozások alapján a GR szempontjából jobb beágyazást eredményezhet (bár a logloss-t rontja).

\end{document}